\documentclass[%
preprint, 10pt,
 amsmath,amssymb,  aps, prd
]{revtex4-2}
\usepackage{slashed}
\usepackage{mathrsfs}

\usepackage{graphicx}
\graphicspath{{pics/}}
\usepackage{dcolumn}
\usepackage{bm}
\usepackage{hyperref}

\usepackage[utf8]{inputenc}
\usepackage{color}
\allowdisplaybreaks[1]

\usepackage[normalem]{ulem}

\begin{document}

\title{Energy-driven disorder in the mean field QCD}

\author{Sergei N. Nedelko}
\email{nedelko@theor.jinr.ru}
\author{Vladimir E. Voronin}%
\email{voronin@theor.jinr.ru}
\affiliation{Bogoliubov Laboratory of Theoretical Physics, JINR, 141980 Dubna, Russia}%

\date{2020}

\begin{abstract}
An impact of the finite size effects on the vacuum free energy density of full QCD with $N_{\rm f}$ massless flavors in the presence of
homogeneous  (anti-)self-dual Abelian background gluon field is studied. The zero temperature free energy density of the four-dimensional spherical domain is computed as a function of the background field strength $B$ and domain radius $R$. Calculation is performed in the one-loop approximation improved by accounting for mixing of the quark and gluon quasi-zero modes with normal modes, with the use of the $\zeta$-function regularization. It is indicated that, under plausible assumption on the character of the mixing, the quantum correction to the free energy density  has a minimum as a function of $B$ and $R$. Within the mean field approach to QCD vacuum  based on domain wall network representation of the mean field, an existence of the minimum may prevent infinite growth of individual domain, thus protecting the vacuum from  the long-range ordering, and, hence, serving as the origin of disorder in the statistical ensemble of domain wall networks, driven by the minimization of the overall free energy of the dominant gauge field configurations.

\end{abstract}

\maketitle

	\section{Introduction}

It is generally accepted that physical QCD vacuum can be characterized by various gluon, quark and mixed condensates.  The condensates have played  an important role in understanding the basic features of hadron physics. In particular, the lowest dimension condensates  $\langle g^2F^2\rangle$, 
$\langle (g^2\tilde F F)^2\rangle$ and $\langle\bar\psi \psi\rangle$ are relevant to the anomalous breakdown of scale and $U_{ \mathrm A}(1)$ symmetries, spontaneous breaking of chiral  $SU_{\mathrm L}(N_f)\times SU_{\mathrm R}(N_f)$ symmetry.  

In principle, the condensates could be  described in terms of the background (vacuum) gluon fields within the self-consistent mean field approach to QCD vacuum. However,  necessity to express 
the condensates, which are vacuum expectation values of the color neutral composite fields,  in terms of the vector potential of the background gauge field in pair with the strong coupling regime complicates the actualization  of the mean field approach to QCD vacuum. 

Depending on base standpoint, the gauge mean field configurations underlying the condensates have been taken in various forms ranging from superposition of quasi-classical gluon configurations like the instanton gas or liquid \cite{Callan:1977gz,Diakonov:1983hh} to the fields with constant field strength squared, $g^2F^2={\rm const}$, representing the global minimum of the quantum effective action of QCD.  Properties of the quantum effective action for   homogeneous gluon  fields  were studied in various approaches~\cite{Eichhorn:2010zc,Savvidy:1977as,Pagels:1978dd,Minkowski:1978fv,Leutwyler:1980ev,Leutwyler:1980ma,Trottier:1992xp, Flory:1983td,Flory:1983dx, Elizalde:1984zv}. In particular, Leutwyler has demonstrated that the covariantly constant Abelian (anti-)self-dual  field 
\begin{eqnarray}
\hat B_\mu=-\frac{1}{2}\hat n B_{\mu\nu}x_\nu, \ \tilde B_{\mu\nu}=\pm B_{\mu\nu} \ B_{\mu\rho}=B^2 \delta_{\nu\rho}, \ \hat n=t_3\cos\xi + t_8 \sin\xi,
\label{bfield}
\end{eqnarray} 
 is singled out from other gluon fields as the only gauge field configuration with constant strength which is stable against small gluon and quark fluctuations and, hence, could be considered as likely contender for the global minimum of the effective action~\cite{Leutwyler:1980ma}. The stability was understood as the absence of the tachyonic modes in the spectrum of small quantum fluctuations in a given background gluon field. An obvious shortcoming of the purely homogeneous gauge field as a candidate for the mean vacuum field is that it would describe a globally ordered vacuum state and thus break all the symmetries of QCD. 
 However, this argument does not apply to the more complicated relevant case when  the gluon fields minimizing quantum effective action, the vacuum fields,  belong to a set of 
lumpy configurations  corresponding to  distributed in $R^4$ domains of homogeneous Abelian (anti-)self-dual gluon field with size and shape randomly varying around certain mean values~\cite{Nedelko:2014sla}. 
As a whole,  such a set  can be characterized as the statistical  ensemble of gauge fields  being  homogeneous Abelian (anti-)self-dual almost everywhere in $R^4$ besides the boundaries between domains where the field appears to be neither homogeneous nor Abelian (anti-)self-dual.   
It has to be noted that topological charge density distribution in the typical gauge field configurations was studied within Lattice QCD with dynamical quarks  with the results supporting  the picture of  entangled space-time regions of sign-alternating topological charge~\cite{deForcrand:2006my,Moran:2008xq,Bruckmann:2011ve}.

An instance  of disordered lumpy configurations is  well seen within the  Ginzburg-Landau (GL) modeling of quantum effective action of QCD~\cite{Nedelko:2014sla,Galilo:2010fn,George:2012sb}.  In this approach almost everywhere homogeneous Abelian (anti-)self-dual gluon field  can be represented as the domain wall networks, arising straightforwardly as soon as the existence of the nonzero scalar gluon condensate $\langle g^2F^2\rangle$ is assumed. 
Domain wall networks come out of the structure of the degenerate discrete global minima of the GL effective potential,  related to each other \textit{via} discrete symmetry transformations   --   CP and Weyl reflections in the root space of color $su(3)$  algebra.

Domain-structured configurations and homogeneous field are almost  indistinguishable with respect to important bulk properties like
\begin{eqnarray*}
 \lim_{V\to\infty}V^{-1}\int\limits_{V}d^4 x g^2F^a_{\mu\nu}F^a_{\mu\nu}=B_\mathrm{vac}^2\not=0,\\
 \lim_{V\to\infty}V^{-1}\int\limits_{V}d^4 x |g^2\tilde F^a_{\mu\nu}F^a_{\mu\nu}|\simeq B_\mathrm{vac}^2.\\
\end{eqnarray*}
For domain wall network with sufficiently thin boundaries between domains,  both gauge field invariants $F^2$ and $\tilde FF=\pm F^2$ are nonzero and constant  almost everywhere in $R^4$.

Almost everywhere homogeneous Abelian (anti-)self-dual gluon fields 
 have been incorporated into hadronization scheme within the domain model of QCD vacuum~\cite{Efimov:1995uz,Burdanov:1996uw,Kalloniatis:2001dw,Kalloniatis:2003sa,Nedelko:2016gdk,Nedelko:2016vpj}.    The Abelian (anti-)self-dual nature   ensures  confinement of both dynamical and static quarks (absence of  poles in the propagators of color charged fields as well as fulfillment of the area law for Wilson loop), resolution of the $U_A(1)$ problem, and  spontaneous breakdown of chiral symmetry. With minimal set of parameters, the domain model provided universal and rather accurate  description of the  masses and various decay constants of light, heavy-light mesons and heavy quarkonia,  including their excited states, as well as some  form factors. The picture of QCD vacuum based on  Abelian (anti-)self-dual  mean field turned out to be suggestive  for exposing a  catalyzing impact of strong electromagnetic field on quark deconfinement~\cite{Galilo:2010fn,
Bali:2011qj,Bali:2014kia,Nedelko:2014sla,Bonati:2016kxj}.  
 
 On the whole, these rather satisfactory  phenomenological applications  put forward a task to  clear up the mechanism  behind the  balance between the competitive tendencies for a long-range order in the ground state  and disorder that may  originate from two complementary origins - the topologically stable defects in the  background field  and existence of a minimum of the effective action with respect to the size of the regions of homogeneity (domain size). 
 Topologically nontrivial gluon field configurations  are expected to emerge through the division of the arbitrary gauge field into Abelian and non-Abelian parts, and may be seen in the domain wall network as frustrations of the color and space-time orientation of the background field at the domain wall junctions. We shortly comment on this mechanism below but do not discuss it in detail here. Discussion of relevant physics  can be found in review~\cite{Kondo:2014sta} and references therein.

Existence of a minimum of the effective action with respect to the domain size can be called the energy-driven origin of disorder, which is the main subject of the present paper. The quark and gluon quasi-zero modes characteristic for the covariantly constant Abelian (anti-)self-dual background field in a finite region  may play   peculiar role in formation of domains.  Infinite number of quasi-zero modes become degenerate zero modes in the limit of  the  infinite domain size and thus potentially lead to an infra-red singular behaviour of the effective potential.   In the plain one-loop approximation, contribution of the quark and gluon quasi-zero modes to the effective potential have opposite signs, and their strong concurrent  impact on the potentially IR singular behavior of the effective potential  becomes manifest in the infinite volume limit.

In the present paper  the effective potential  for the Abelian (anti-)self-dual field \eqref{bfield} in a four-dimensional spherical domain with radius $R$ is calculated  for $SU(3)$ chromodynamics with $N_{\rm f}$ massless quarks using the zeta function regularization, which completes the previously reported results for pure gluodynamics~\cite{Nedelko:2019fao}. The quark and gluon fields are subject to the bag-like boundary conditions~\cite{Kalloniatis:2001dw,Nedelko:2014sla}. 
It is important that interaction of gluon and quark quasi-zero modes ( zero modes in the limit $R\to\infty$) with the normal modes has to be treated beyond one loop as it has been  put forward by Leutwyler~\cite{Leutwyler:1980ma}. This interaction leads to the  contribution of the quasi-zero modes to the effective potential being regular in the  limit of infinite domain size. This  is crucial for overall existence of the thermodynamic limit and, as underlined in~\cite{Leutwyler:1980ma},   consistency of the strong field limit of the  effective action  with the asymptotic freedom.  In fact, a mixing of quasi-zero and normal modes has to be taken into account in all-loop orders as there is no good small parameter, which presents a hardly solvable task especially in the finite region. However, as it was demonstrated in~\cite{Leutwyler:1980ma}, the most significant consequence of the mixing is the emergence of the effective "mass" for the quasi-zero modes. This observation enables sensible  modeling of the plausible form of dependence of the  "effective mass" on the domain size $R$, and identification of conditions for the existence of the  minimum of the effective potential with respect to the radius of the domain $R$ and field strength $B$ inside domain. 

It is shown that formation of the domains with finite size $R$ can be energetically preferable if the effective "mass" falls from asymptotic nonzero value at $R\to\infty$, fixed by the asymptotic freedom and 
correct strong field limit, to zero at $R\to 0$ as it follows from 
the dependence of normal modes on the domain size. The final result for the free energy density of full QCD is shown in 
Fig.~\ref{effpot_total_figure}. The minimum in  field strength and domain size is clearly present for a wide class of functional dependence of the zero mode effective "mass" on the domain size.

Finite mean size of the domains in the network is determined by the minimum of the effective action density inside individual domains.  Existence of the mean domain size minimizing action density can be seen as a condition for sustainability of a domain wall network. In general,  there is an infinite set of networks with degenerate values of quantum effective action that constitutes a statistical ensemble of dominant vacuum gluon field configurations.  In accordance with the character of free energy dependence on domain size, the degree of disorder in the ensemble may vary from highly disordered
distribution of entangled domains with variable size and shape to the almost periodic distribution of identically sized and shaped domains, reminiscent of spin liquid and (anti-) ferromagnetic state respectively.

In the next section we update the formal framework underlying the domain wall network representation of the background  gluon fields  in order to adjust it to the formulation of the  particular problem studied in this paper. The ghost, gluon and and quark contributions to the  free energy density of a spherical domain  with Abelian self-dual gauge field   are discussed in detail in the third section.   Appendices contain quite involved technical details of calculations, which are the result of the present paper in themselves. 

\section{Effective action of QCD and the domain wall networks}

The initial, gauge unfixed, Euclidean functional integral representation for  QCD partition function 
\begin{eqnarray*}
 &&Z[B_{\rm vac}]=N\int\limits_{{\cal F}} DA\int \limits_{\Psi} D\bar\psi D\psi \exp\{-S[A,\bar\psi,\psi]\},
 \end{eqnarray*}
assumes certain choice of the functional spaces of integration over gluon and quark fields.  If one allows nonzero gluon condensates then the functional space ${\cal F}$ has to be  subjected to an appropriate condition, for instance 
 \begin{eqnarray}
 \label{cond0}
 {\cal F}=\left\{A: \lim_{V\to \infty} \frac{1}{V}\int\limits_V d^4xg^2F^a_{\mu\nu} (x)F^a_{\mu\nu}(x) =B_{\rm vac}^2\right\}.
\end{eqnarray}
This definition is reminiscent of the Schrödinger functional representation (see for instance~\cite{Leutwyler:1980ma,Faddeev:2009cd,Faddeev:2015sda}. The difference is that condition  \eqref{cond0}
is imposed onto the  gauge invariant combination of the gauge fields and has an integral (functional) form. 
 Division
of the general gauge fields $A_\mu^a$ into the background fields $B_\mu^a$ with extensive classical action specified by Eq.~\eqref{cond0} and the fluctuations $Q_\mu^a$ in the background $B_\mu^a$  supplemented by the background gauge condition  $D(B)Q=0$  leads to the representation~\cite{Galilo:2011nh}
\begin{eqnarray}
Z[B_{\rm vac}] &=&N'\int\limits_{{\cal B}}DB \exp\left\{-S^{V}_{\rm eff}[B]\right\} 
\\&=&N''\int\limits_{{\cal B}}DB \int\limits_{{\cal Q}} DQ 
\int \limits_{\Psi} D\bar\psi D\psi\det[D(B)D(B+Q)]
\delta[D(B)Q] \exp\{-S_V[B+Q,\bar\psi,\psi]\}.
\nonumber
\label{zqcd2}
\end{eqnarray}
At this step, functional spaces ${\cal Q}$ and $\Psi$ are restricted by the condition
\begin{eqnarray}
\lim_{V\to\infty}\left( S_V[B+Q,\bar\psi,\psi]-S_V[B]\right)<\infty,
\label{condqcd2}
\end{eqnarray}
which  excludes long-range fields from the set of fluctuations.
Integral over the quark $\psi$ and gluon $Q$ fluctuations   defines effective action $S^V_{\rm eff}[B]$ for a given background field $B$.  
Whether the constant $B_{\rm vac}$ is nonzero has to be determined by the minima of the quantum effective action.
In the infinite volume limit $V\to\infty$ the global minima
 of  $S^V_{\rm eff}[B]$ dominate the integral over background fields $B$ and thus determine the specific class of  gluon field configurations relevant to the self-consistent mean field description of QCD vacuum. 
 
Fields with a constant field strength have to be verified for the role of the vacuum mean field $B_\mu^a$ first. Such a verification has been going since late seventies when chromomagnetic covariantly constant Abelian gauge field has been suggested for the role of QCD vacuum~\cite{Savvidy:1977as}.  As has already been mentioned, covariantly constant Abelian (anti-)self-dual gauge field has appeared to be more preferable sample in many aspects, in particular due to the direct relation to confinement of dynamical color charged fields, chiral symmetry breaking, seen also in terms of meson properties through hadronization.    

It should be stressed that condition \eqref{cond0} in no way restricts the background field functional space ${\cal B}$ to configurations with constant strength. Representation \eqref{zqcd2} assumes  division of arbitrary gauge field $A$ into two parts, the  background $B$ and fluctuation $Q$ fields with simultaneous gauge fixing for the fluctuation part.  If one is going to study Abelian background $B$ then a specific parametrization of the gauge field  has to be implemented~\cite{Shabanov:1999mu, Shabanov:2000dn,Cho:1980vs,
Faddeev:2006sw, Kondo:2005eq,Kondo:2014sta}). 
Both color and space orientations of the background field may appear to be frustrated at some space-time locations thus making manifest topological singularities in the vector potential, which, in general, cover the whole range of defects of various dimensions -- domain wall, vortex, monopole and zero-dimensional instanton-like defects.  

More importantly, as it has been stressed, in particular,  by Faddeev, quantum equations 
\begin{eqnarray*}
\frac{\delta S^{V}_{\rm eff}[B]}{\delta B}=0
\end{eqnarray*}
"could have soliton solutions, which are absent in the classical limit.  In particular, it is not completely crazy idea that quantum Yang-Mills equations have soliton-like solutions due to the dimensional transmutation"~\cite{Faddeev:2010fa}.

We have to conclude that 
representation of the dominant vacuum gauge field configurations  by the plain covariantly constant field, corresponding to a globally ordered ground state, is an  extreme case.
Non-homogeneous at certain space-time locations   field configurations with extensive classical action   may have lower effective action than constant fields or can be topologically protected  and should be taken into consideration anyway.    

In general, fields which dominate the functional integral in the infinite volume (thermodynamic) limit belong to functional subspace  $\tilde{\cal B}\subset {\cal B}$. Given this,  in the infinite volume limit one may reduce the  integration over background fields $B$ in \eqref{zqcd2} to the subspace of dominant (vacuum) fields and arrive at the mean field representation of the QCD partition function
\begin{eqnarray}
Z[B_{\rm vac}] =N''\lim_{V\to\infty}\int\limits_{\tilde{\cal B}}D\sigma_B \int\limits_{{\cal Q}} DQ 
\int \limits_{\Psi} D\bar\psi D\psi \det[D(B)D(B+Q)]
\delta[D(B)Q] 
\nonumber
\\
\times\exp\{-S_V[B+Q,\bar\psi,\psi]\},
\label{zqcd3}
\end{eqnarray}
where $D\sigma_B$ is a measure of integration over the space of vacuum gauge fields.
A treatment of these possibly soliton-like vacuum fields $B\in \bar{\cal B}$ in the functional integral \eqref{zqcd2} must be nonperturbative, while 
perturbative expansion over fluctuations is likely to be applicable  for calculation of various physical quantities~\cite{Nedelko:2016gdk,Nedelko:2016vpj}.  It has to be noted that according to condition (\ref{zqcd2}) normalization constant $N''$ should contain a factor which cancels a trivial extensive contribution of the background field to the classical action $S[B]$ in the infinite volume limit.    Certain prescriptions for regularization and renormalization of UV divergences are assumed. Due to the dimensional transmutation in gauge theories the condensates have to be expressed in terms of internal scale $\Lambda_{\rm QCD}$, that is $B_{\rm vac}\propto \Lambda^2_{\rm QCD}$.  
Interrelation between dimensional transmutation in gauge theories and gauge field condensates was discussed in~\cite{Faddeev:2006bm}.

The properties of the effective action  are crucial for
 practical implementation of the described scheme. Details of discussion of the very existence of nontrivial minimum of the quantum effective action for homogeneous fields  can be found in papers~\cite{Eichhorn:2010zc,Savvidy:1977as,Pagels:1978dd,Minkowski:1978fv,Leutwyler:1980ev,Leutwyler:1980ma,Trottier:1992xp, Flory:1983td,Flory:1983dx, Elizalde:1984zv}. 
Some basic properties have been estimated and can be used for identification of the likely features of the mean field subspace  
$\tilde{\cal B}$.

An important observation was that  in the infinite volume limit calculation of the effective potential for covariantly constant gauge fields  could not be performed within the plain one-loop approximation because of  the presence of infinitely degenerate tachyonic (for chromomagnetic field) or zero modes (for (anti-)self-dual field). One has to improve the one-loop calculation by accounting for  interaction of the   zero modes with normal modes~\cite{Leutwyler:1980ma} in order to  generate a well-defined Gaussian measure for zero modes. The issues of tachyonic and zero modes were addressed in papers~\cite{Flory:1983td,Flory:1983dx, Elizalde:1984zv} on the basis of specific scaling properties of the massless QCD, concluding that tachyonic and zero modes  seem to be the artifacts of the one-loop approximation. Functional renormalization group calculation of the effective potential for Abelian (anti-)self-dual covariantly constant field~\cite{Eichhorn:2010zc} does not encounter any problem with gluon zero modes. It has confirmed earlier estimates based on the improved one-loop approximation for the effective potential \cite{Leutwyler:1980ma} and existence of a  minimum at nonzero homogeneous Abelian (anti-)self-dual gluon field.

Ginzburg-Landau approach to the quantum effective action of QCD with effective Lagrangian~\cite{Kalloniatis:2001dw,Galilo:2010fn,Nedelko:2014sla,George:2012sb}
\begin{eqnarray}
\mathcal{L}_{\mathrm{eff}} = - \frac{1}{4\Lambda^2}\left(D^{ab}_\nu F^b_{\rho\mu} D^{ac}_\nu F^c_{\rho\mu} + D^{ab}_\mu F^b_{\mu\nu} D^{ac}_\rho F^c_{\rho\nu }\right)  -U_{\mathrm{eff}},   
\nonumber \\
U_{\mathrm{eff}}=\frac{\Lambda^4}{12} \mathrm{Tr}\left(C_1\breve{ f}^2 + \frac{4}{3}C_2\breve{ f}^4 - \frac{16}{9}C_3\breve{ f}^6\right) \label{effpot}
\end{eqnarray}
 indicated an   intrinsic possibility for disordered ground state of QCD.
Here $\Lambda$ is a scale,  $F^a_{\mu\nu}$ is the standard strength tensor for 
$ SU_{\rm c}(3)$ color gauge field,  $\breve f_{\mu\nu}= T^a F^a_{\mu\nu}/\Lambda^2$,  $D^{ab}_\mu = \delta^{ab} \partial_\mu - i\breve{ A}^{ab}_\mu$. 
The effective Lagrangian respects all symmetries of QCD besides scale invariance, and the real constants $C_i$ have to be positive to provide a minimum of the effective potential at nonzero gauge field. 
Given this, one can check that there is a discrete set of global  minima  corresponding to the covariantly constant Abelian (anti-)self-dual fields
\begin{equation*}
 \breve A^{k}_{\mu}  = -\frac{1}{2}\breve n_k F_{\mu\nu}x_\nu, \, \tilde F_{\mu\nu}=\pm F_{\mu\nu},\quad F_{\mu\nu}F_{\mu\nu}=b^2\Lambda^4,\quad b_\mathrm{vac}^2=\frac{-C_2+\sqrt{C_2^2+3C_1C_3}}{3C_3},
\end{equation*}
where the matrix $\breve{n}_k$ belongs to the Cartan subalgebra of $su(3)$
\begin{equation}
\label{n_specific}
\breve n_k = T^3\ \cos\left(\xi_k\right) + T^8\ \sin\left(\xi_k\right),
\quad
\xi_k=\frac{2k+1}{6}\pi, \, k=0,1,\dots,5.
\end{equation}
These minima are connected with each other  by discrete  parity and Weyl symmetry transformations. 

One concludes that the domain wall solutions of the quantum equations of motion emerge as soon as the effective action has a global minimum  corresponding to nonzero  gluon condensate $\langle g^2F^2\rangle$.
For instance, if all parameters of the field besides angle $\omega$ between chromoelectric and chromomagnetic fields are put to the vacuum values then initial GL Lagrangian describes a sine-Gordon field $\omega$, 
\begin{equation*} 
\mathcal{L}_{\mathrm{eff}} = -\frac{1 }{2}\Lambda^2 b_{\mathrm{vac}}^2 \partial_\mu \omega \partial_\mu \omega 
- b_{\mathrm{vac}}^4  \Lambda^4 \left(C_2+3C_3b_{\mathrm{vac}}^2 \right){\sin^2\omega}.
\end{equation*}
 The available standard kink solution describes a planar domain wall between the regions with homogeneous Abelian self-dual and anti-self-dual gluon fields. Topological charge density vanishes on the wall where the chromomagnetic and chromoelectric fields are orthogonal to each other.

\begin{figure}
\includegraphics[width=0.5\textwidth]{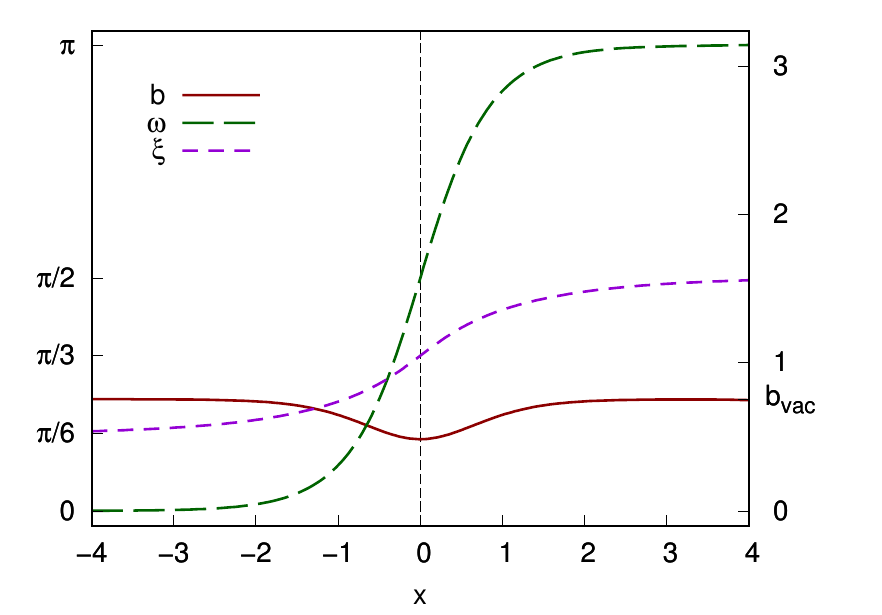}
\caption{Domain wall solution of Eqs.~\eqref{3-eq_xi}: gauge field interpolates between two vacuum states, namely self-dual and anti-self-dual configurations and $\xi$  going from one boundary of Weyl chamber to another, field strength $b$ has a dip inside the wall.
\label{3-eq_solution2}}
\end{figure}
\begin{figure}
	\hspace*{0.1\textwidth}
	\includegraphics[width=0.2\textwidth]{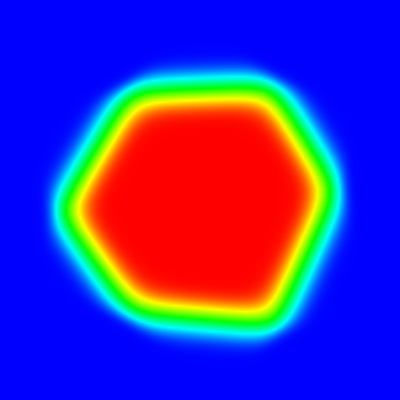}
	\hfill
		\includegraphics[width=0.2\textwidth]{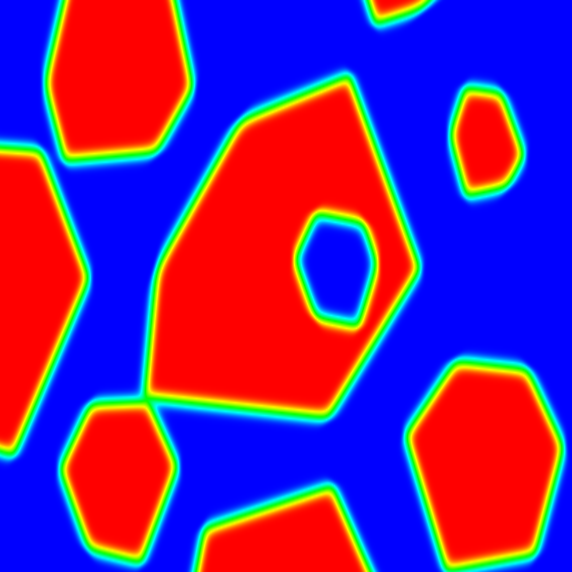}
	\hfill
	\includegraphics[width=0.2\textwidth]{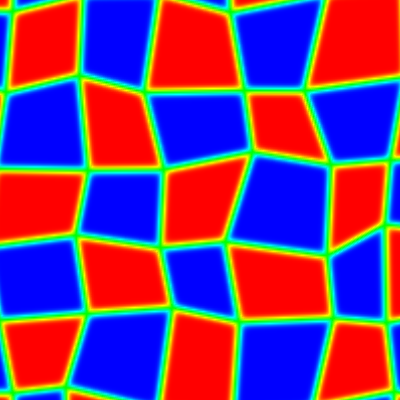}
	\hspace*{0.1\textwidth}\mbox{}\\
	\caption{Two-dimensional slices of topological charge density for various superpositions of domain wall solutions for Eqs.\eqref{3-eq_xi}. Red and blue colors correspond to Abelian self-dual and anti-self-dual field.
	\label{sample_net}}
\end{figure}

More general domain wall correspond to $\omega$, $\xi$ and $b$ varying simultaneously. In this case equations of motion read:
\begin{eqnarray}
&&-6 b' \omega'+b^3 \sin 2 \omega 
\left(C_3 b^2 (\cos 6\xi+10)+3
C_2\right)-3 b \omega''=0,
\nonumber
\label{3-eq_omega}
\\
&&-15 b''+3 b
\left(-4 C_1+2 \omega'^2+5 \xi'^2\right)-12
C_2 b^3 (\cos 2 \omega -3)-2
C_3 b^5 (3 \cos 2 \omega -5) (\cos
6\xi+10)=0,
\nonumber
\label{3-eq_b}
\\
&& 2 C_3 b^6 (3 \cos 2 \omega-5) \sin 6\xi
-15b \left(2 b' \xi'+b \xi''\right)=0.
\label{3-eq_xi}
\end{eqnarray}
The plain domain wall solution of these equations is shown in Fig.~\ref{3-eq_solution2}.
It corresponds to gauge field which interpolates between two different vacuum configurations. In general, initial GL Lagrangian may produce more nontrivial soliton-like solutions both in Euclidean and Minkowski metrics. 
A combination of the additive and multiplicative superpositions of the domain walls  allow one to generate various domain walls and domain wall networks in $R^4$~\cite{Vachaspati:2006gu,Nedelko:2014sla}, like samples shown in  
Fig.~\ref{sample_net}.

Lagrangian \eqref{effpot} has the simplest form, but its symmetry properties and  emergence of the  periodic discrete  minima as a consequence of the scale invariance breakdown  seem to be a general property, qualitatively insensitive to the detailed form of the effective potential.  Another form of the strong field behavior of the GL effective Lagrangian can affect the particulars of the kink solution but can hardly influence its very existence and general properties.  It has to be noted that the role of Weyl reflections in topology of QCD ground state has been intensively discussed in recent years in the context of dual superconductor picture of confinement~\cite{Cho2,Kondo:2014sta}.

  The functional space  $\tilde{\mathcal{B}}$ in the integral \eqref{zqcd3} is assumed to include infinitely many networks with equal values of the free energy.   Though implicit implementation of this prescription within the domain model of QCD vacuum demonstrated high phenomenological 
performance~\cite{Nedelko:2016gdk,Nedelko:2016vpj}, the conceptual problem of the network stability remains: since domain wall configurations in four-dimensional Euclidean space are not topologically protected and  the presence of kink configuration usually increases action then any network should evolve to a single infinitely large domain. In the next section we study a particular effect which may prevent an infinite growth of a single domain.

\section{Free energy density}

Lagrangian \eqref{effpot} does not account for possible finite size effects.   Available  evaluations of the effective potential for homogeneous gauge field which were performed in the infinite space-time~\cite{Eichhorn:2010zc,Savvidy:1977as,Pagels:1978dd,Minkowski:1978fv,Leutwyler:1980ev,Leutwyler:1980ma,Trottier:1992xp, Flory:1983td,Flory:1983dx, Elizalde:1984zv}, thus,  as a matter of fact,   implicitly assuming that the free energy density  in the region with homogeneous field does not depend on the size of the region. 
  Meanwhile, finite size effects are not excluded and may prevent infinite growth of an individual domain,  thus protecting overall stability of the domain wall network configurations. 
  
  The present section is devoted to the study of the dependence of the renormalized free energy density $F(B,R)$ of a finite spherical domain of the Abelian (anti-)self-dual homogeneous field on its radius $R$ and the strength $B$ of the field in full QCD with massless quarks, defined by the finite volume partition function
 \begin{eqnarray}
 \exp(-V_RF(B,R))={\rm Ren} \ N\int\limits_{{\cal Q}} DQ 
\int \limits_{\Psi} D\bar\psi D\psi\int \limits_{\cal C} D\bar c D c 
\ \exp\left\{-S_{V_R}[Q;\bar\psi,\psi;\bar c, c;B]\right\},
\label{intF}
\end{eqnarray}
where $V_R$ is the volume of four-dimensional ball with radius $R$, which is an idealization of the domain shown in the leftmost picture in Fig.\ref{sample_net}, and $S_{V_R}$ is the  gauge-fixed action of QCD in the presence of the background gluon field defined by Eq.\eqref{bfield}. The  Feynman background gauge is used below.
Normalization constant $N$ is fixed by the condition
\begin{eqnarray}
F(0,R)=0.
\label{NF}
\end{eqnarray}
Renormalization prescription is  specified below.
Functional spaces $\cal{Q}$, $\Psi$ and $\cal{C}$ contain 
the quark, gauge and ghost fields subject to the bag-like boundary conditions 
\begin{eqnarray}
\label{bag_bc}
&&\left.\left(i \slashed{\eta}e^{i\alpha\gamma_5}- 1\right)\psi(x)\right|_{x\in \partial V_R}=0,
\\
&&\label{dirichlet_bc}
\left. \breve  n Q_\mu(x)\right|_{x\in \partial V_R}=0, 
\\
\label{dirichlet_bcc}
&&\left. \breve n c(x)\right|_{x\in \partial V_R}=0. 
\end{eqnarray}
These boundary conditions were discussed in papers~\cite{Nedelko:2014sla,Kalloniatis:2001dw}. The choice assumes that there is a physical boundary of the spherical domain, given by the domain wall illustrated in Fig.\ref{3-eq_solution2}. Bag-like boundary conditions are required by the qualitatively different character of field fluctuations in the bulk of domain (confining self-dual background field) and on the boundary (chromomagnetic field)~\cite{Nedelko:2014sla}. 

The functional integral \eqref{intF} is defined through decomposition of the quark, gauge and ghost fields,
\begin{equation*}
Q(x)=\sum_{n}q_n Q^{(n)}(x),
\quad
c(x)=\sum_{n} c_n C_n(x),
\quad
\psi(x)=\sum_n \theta_n \Psi_n(x),
\end{equation*}
over  orthogonal normalized complete set functions in $\Psi$, $\cal{Q}$ and $\cal{C}$. It is convenient to  diagonalize quadratic part of the action, using the eigenfunctions of the corresponding differential operators,
\begin{eqnarray}
&&\hat{\slashed{D}}\Psi_n=\lambda_n^\text{q}\Psi_n,
\label{eignvpq}
\\
&&-\breve D^2 C_n = \lambda_n^\text{gh} C_n,
\label{eignvpgh}
\\
&&\left[-\breve D^2\delta_{\mu\nu}+2i\breve B_{\mu\nu}\right]Q^{(n)}_\nu=\lambda_n^\text{gl} Q^{(n)}_\mu,
\label{eignvpgl}
\end{eqnarray} 
subject to boundary conditions~\eqref{bag_bc}, \eqref{dirichlet_bc} and \eqref{dirichlet_bcc}.
Indices $n$  denote all relevant quantum numbers as described below. The quark, gluon and ghost spectra 
 are purely discrete for any $R$ if field strength $B$ is nonzero. 
At finite $R$, all eigenvalues are nonzero for quark fields, and positive  for gauge and ghost fields.
The one-loop  correction $\delta U$ to the classical action
\begin{equation*}
U_\textrm{cl}=\int_{V_R} d^4x \frac{1}{4g^2}\left(B^a_{\mu\nu}\right)^2=\frac{\pi^2B^2R^4}{2g^2}.
\end{equation*}
 is  given by the determinants:
\begin{equation}
\label{one-loop_via_det}
\exp\left(-\delta U\right)=N\left[\det \left(-\breve D^2\delta_{\mu\nu}+2i\breve B_{\mu\nu}\right)\right]^{-\frac{1}{2}}\ \det \left(-\breve D^2\right)\ \left[\det i\slashed{D}\right]^{N_f},
\end{equation}
where $N_f$ is the number of massless quark flavors. 
Renormalized functional determinants are calculated below by means of analytical regularization
\begin{equation*}
\mathrm{Tr}\log\Delta=-\frac{d}{ds}\sum_n \left. \lambda_n^{-s}\vphantom{\frac11}\right|_{s=0}=-\left.\frac{d}{ds}\zeta(s)\right|_{s=0},
\end{equation*}
where $\lambda_n$ are eigenvalues of operator $\Delta$.
Computation of $\zeta(s)$ is based on the method summarized in  papers~\cite{Bordag:1995gm,Kirsten:2001}.

\subsection{Ghost contribution to the free energy density}

Evaluation of the ghost contribution to the free energy density is rather straightforward.
Details of solution of the eigenvalue problem, Eqs.\eqref{eignvpgh} and \eqref{dirichlet_bcc}, are given in paper~\cite{Kalloniatis:2001dw}. Ghost eigenfunctions are expressed in terms of confluent hypergeometric function $M(a,b,z)$.
The eigenvalues are defined by equation \eqref{eignvpgh},  which can be written in the form
\begin{equation}
\label{eigenghB}
M\left(\frac{k}{2}+1-m-\frac{\lambda^2}{2v_a B},k+2,\frac{v_a B R^2}{2}\right)=0, k=0,1,2,\dots,\ m=-\frac{k}{2},-\frac{k}{2}+1,\dots,\frac{k}{2},
\end{equation}
where $v_a$ is the absolute value of the $a$-th nonzero eigenvalue of the matrix $\breve{n}=n^aT^a$, and $T^a$ are generators in the adjoint representation. The eigenmodes corresponding to the zero eigenvalues of 
$\breve{n}$  do not contribute to the free energy density $F(B,R)$  due to the normalization condition \eqref{NF}. Eigenvalues with given $k,m$, radial number $r$ and color index $a$ are $(k+1)$-degenerate. 
Ghost contribution takes the form
\begin{equation}
\label{Ugh1}
\delta U^\mathrm{gh}=-\sum_{kmr}\mathrm{Tr}\ln\frac{\lambda_{kmr}^2(vB,R)}{\lambda_{kmr}^2(0,R)}=\left.\frac{d}{ds}\zeta_\mathrm{gh}(s)\right|_{s=0},
\end{equation}
where $\mathrm{Tr}$ denotes summation over  $v^a$.   Dimensionless quantities 
\begin{equation}
\lambda = \lambda/\mu,\quad B = B/\mu^2,\quad R = R\mu
\label{muscale}
\end{equation}
have been introduced using auxiliary renormalization scale $\mu$. 
In the limit $B\to 0$ Eq.~\eqref{eigenghB}   transforms to (see Appendix~\ref{appendix_Kummer_functions})
\begin{equation*}
(k+1)!\left(\frac{\lambda R}{2}\right)^{-k-1}J_{k+1}(\lambda R)=0,
\end{equation*}
which defines eigenvalues $\lambda(0,R)$ in Eq.~\eqref{Ugh1}. 
Zeta function can be written in the 
form~\cite{Bordag:1995gm,Kirsten:2001} 
\begin{eqnarray*}
&&\zeta^\text{gh}(s)=\mathrm{Tr}\frac{\sin \pi s}{\pi}\sum_{k=0}^\infty \sum_{m=-\frac{k}{2}}^\frac{k}{2}(k+1) \int_0^\infty \frac{dt}{t^{2s}}\frac{d}{dt}\Psi_\mathrm{gh}(k+1,m,t,\breve{B},R),\\
&&\Psi_\mathrm{gh}(k,m,t,\breve{B},R)=\log\frac{\exp\left(-\frac{\breve{B}R^2}{4}\right)M\left(\frac{k+1}{2}-m+\frac{t^2}{2\breve{B}},k+1,\frac{\breve{B}R^2}{2}\right)}{k!\left(\frac{t R}{2}\right)^{-k}I_{k}(t R)}.
\end{eqnarray*}
This expression is still not suitable for analytical continuation  to  $s\to0$, since intervals of convergence  of the integral and sum do not overlap. To make it ready for continuation to $s\to 0$, several terms of asymptotic series in $k$ of the integrand are added and subtracted:
\begin{eqnarray}
\nonumber
\zeta^\text{gh}(s)&=&
\mathrm{Tr}\frac{\sin\pi s}{\pi}\sum_{k=1}^\infty k^{1-2s} \int_0^\infty \frac{dt}{t^{2s}}\frac{d}{dt}\left[\sum_{m=-\frac{k-1}{2}}^\frac{k-1}{2}\Psi_\mathrm{gh}(k,m,kt ,\breve{B},R)-\sum_{i=0}^2  \frac{u_i^{\mathrm{gh}}(t,\breve{B},R)}{k^i}\right]
\\
\label{ghosts_zeta_start}
&&+\mathrm{Tr}\frac{\sin\pi s}{\pi}\sum_{k=1}^\infty k^{1-2s} \int_0^\infty \frac{dt}{t^{2s}}\frac{d}{dt}\sum_{i=0}^2 \frac{u_i^{\mathrm{gh}}(t,\breve{B},R)}{k^i}.
\end{eqnarray}
The first term  is analytical at $s\to 0$. The sums and integrals in the second term are expressed via analytical functions in the regions of $s$ where they converge, and analytically continued to $s\to 0$ in the complex plane (see Appendix~\ref{appendix_ghost_gluon} for details).
The final expression for $\delta U^\mathrm{gh}$ looks as
\begin{eqnarray}
\nonumber
\delta U^\mathrm{gh}(B,R)&=&-4\sum_{k=1}^\infty k \left[\sum_{m=-\frac{k-1}{2}}^\frac{k-1}{2}\Psi_\text{gh}\left(k,m,0,\frac{\sqrt{3}B}{2},R\right)-\frac{3}{4}B^2R^4\frac{1}{48}\left(1-\frac{1}{k}+\frac{1}{k^2}\right)\right]\\
\label{ghosts_effpot}
&&+\frac{B^2R^4}{48}(2-3\gamma+3\log 2-3\log R)-\frac{3B^4R^8}{10240}.
\end{eqnarray}
The first term is a convergent sum,  ready for numerical computation.

\begin{figure}
\includegraphics[scale=1]{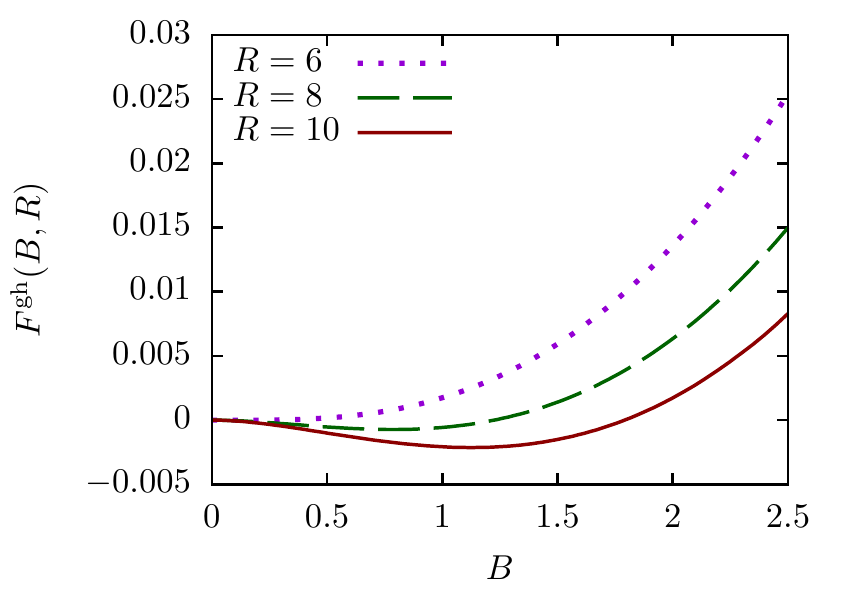}\hfill
\includegraphics[scale=1]{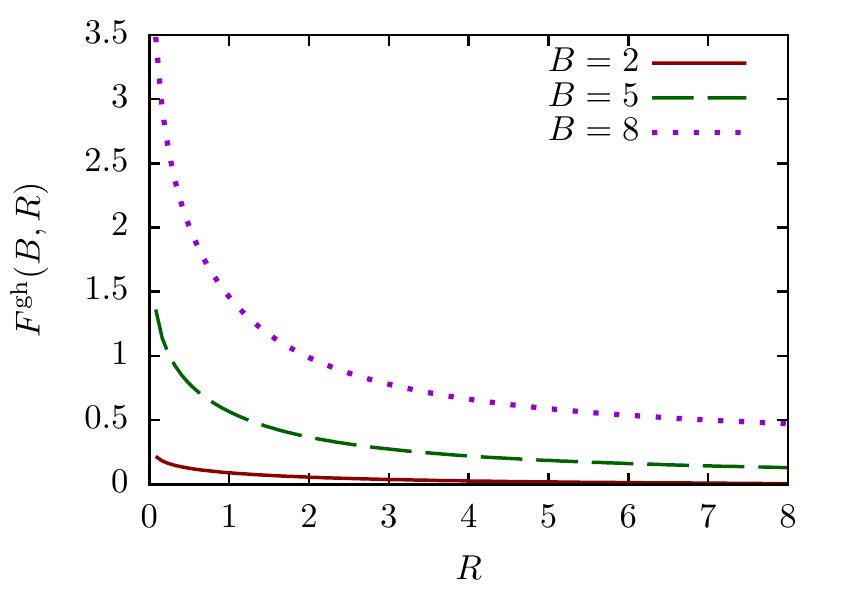}
\caption{Contribution of the Faddeev-Popov determinant to the free energy density.  Domain size $R$ and field strength $B$ are given in units of renormalization scale $\mu$, Eq.~\eqref{muscale}. If $R$ is sufficiently large, then a minimum  exists at nonzero field strength $B$. For any $B$ the ghost contribution is minimal  in the infinite volume limit $R\to\infty$. 
 \label{ghosts_figure}}
\end{figure}

The ghost contribution  $F_\mathrm{gh}(B,R)=\delta U_\mathrm{gh}(B,R)/V_R$ to the free energy density  defined by Eq.~\eqref{ghosts_effpot} is shown in Fig.~\ref{ghosts_figure}.  It demonstrates expected behaviour both in the field strength $B$ and domain size $R$. In the infinite volume limit it approaches expected one-loop ghost contribution.

\subsection{Gauge field fluctuations}

Eigenvalues of relevant gluon operator
 \begin{equation}
-\breve{D}^2\delta_{\mu\nu}+2i\breve{B}_{\mu\nu}
\label{glop}
\end{equation}
differ from the the ghost eigenvalues just by an overall shifts due to the term $2i\breve{B}_{\mu\nu}$.  Though in the infinite space-time this shift leads to the presence of the infinitely many exact gluon zero modes, for  $R<\infty$ the degeneracy of these modes disappears and all corresponding eigenvalues become positive. 

This subset of modes will be referred below as quasi-zero modes. Since exact zero modes are absent for $R<\infty$, then, at first glance, one may hope to compute  gluon contribution straightaway with the result that is usable at least for not very large values of the domain size $R$.
The calculation goes completely analogously  to the computation performed in the previous subsection, and leads to $\delta U^\mathrm{gl}$:
\begin{eqnarray}
\label{gluons_effpot}
\delta U^\mathrm{gl}(B,R)&=&4\sum_{k=1}^\infty k \left[\sum_{m=-\frac{k-1}{2}}^\frac{k-1}{2}\Psi_\text{gl}\left(k,m,0,\frac{\sqrt{3}B}{2},R\right)-\frac{3}{4}B^2R^4\frac{1}{24}\left(1-\frac{1}{k}-\frac{5}{k^2}\right)\right]
\nonumber\\
&-&\frac{B^2R^4}{48}(31+30\gamma-30\log 2+30\log R)+\frac{3B^4R^8}{5120}.
\end{eqnarray}

\begin{figure}
\includegraphics[scale=1]{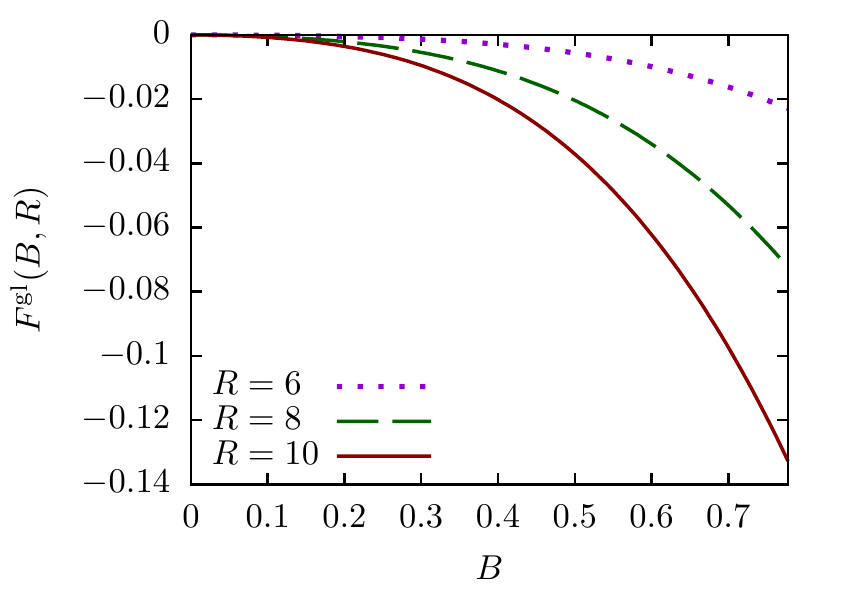}\hfill
\includegraphics[scale=1]{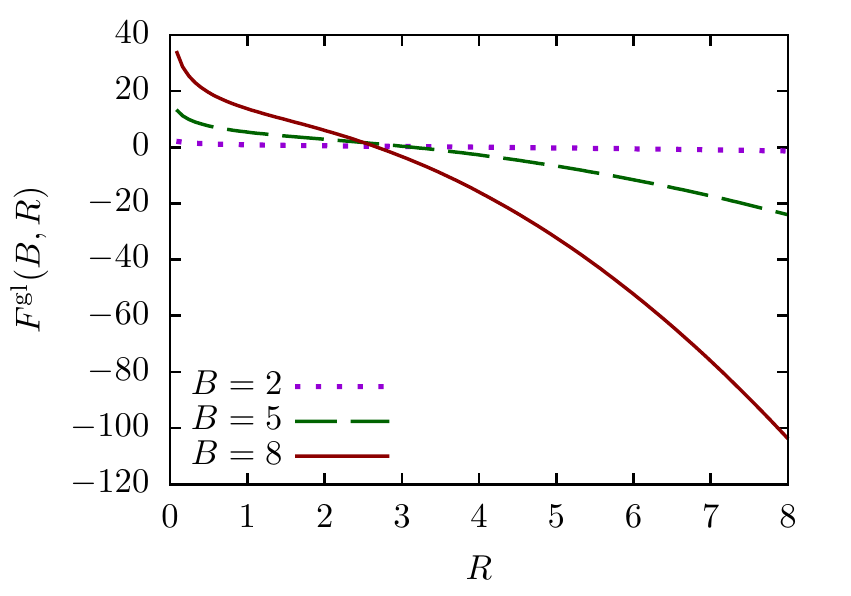}
\caption{Dependence of one-loop gluon contribution to the free energy density on $R$ and $B$ (in units of renormalization scale $\mu$, Eq.~\eqref{muscale}). The strong field and large size volume limits are incorrect due to the presence of infinitely many gluon quasi-zero modes. 
\label{gluon_figure}}
\end{figure}

\begin{figure}
\includegraphics[width=0.49\textwidth]{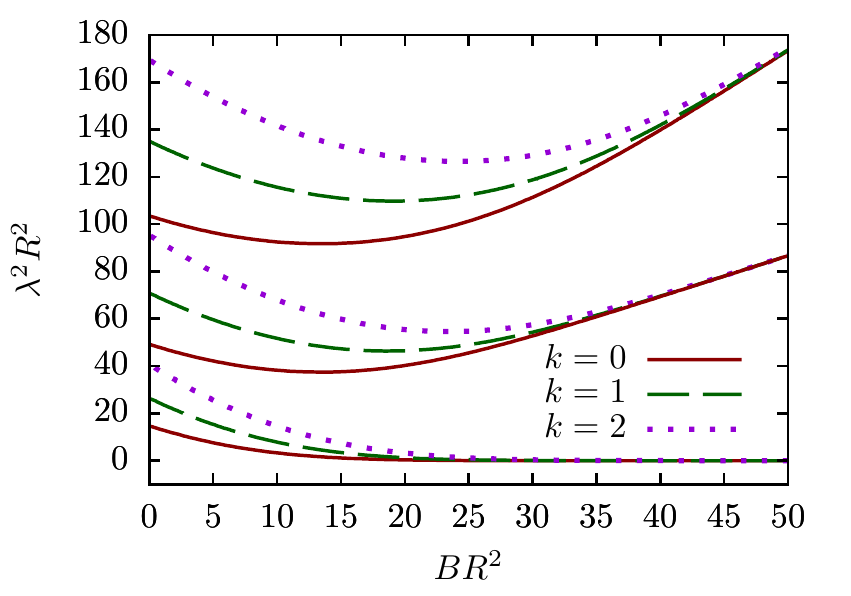}
\caption{Dependence of gluon eigenvalues  (see Eq.~\eqref{gluon_quasizero_modes_equation}) on $BR^2$. In the limit  $BR^2\to \infty$, infinite number of zero modes emerges.\label{figure_lambdas}}
\end{figure}

Free energy density, as it comes out of Eq.~\eqref{gluons_effpot},  is shown in Fig.~\ref{gluon_figure}. One would expect that the character of the strong field limit should be insensitive to the presence of the boundary, since it corresponds to the short distances.  Meanwhile, free energy density decreases without bound at large $B$ and fixed $R$, which does not comply with known results~\cite{Pagels:1978dd, Leutwyler:1980ma}, and, more generally, with asymptotic freedom. 
  It also  does not approach a constant at large $R$ and fixed $B$, that means the absence of sensible thermodynamic limit in the system.  
  This behavior is due to the manifestation of infinitely many quasi-zero eigenvalues that tend to zero as dimensionless quantity $BR^2\to\infty$ (see Fig.~\ref{figure_lambdas}), which occurs both in the strong field and  thermodynamic limits.

\begin{figure}
\includegraphics[scale=0.4]{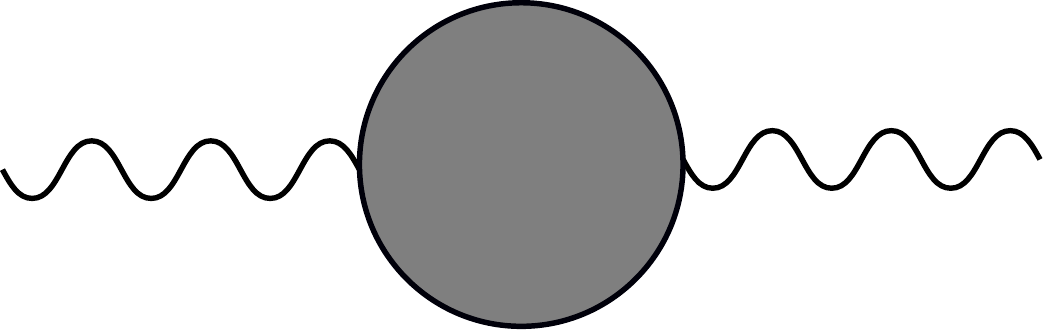}
\hspace{1em}
\includegraphics[scale=0.4]{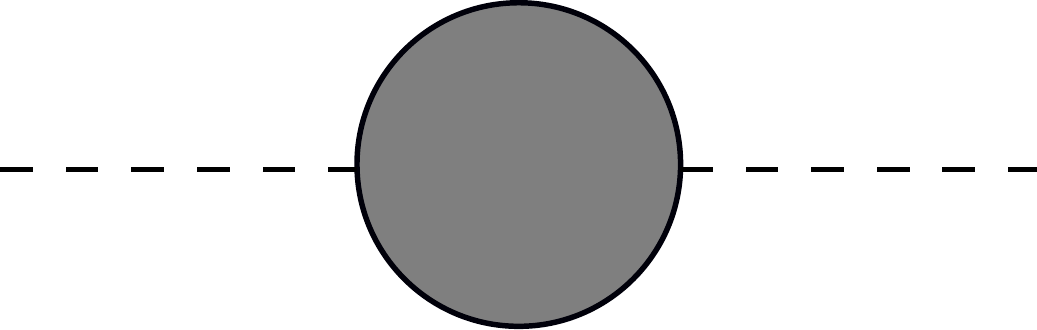}
\caption{ Wavy line corresponds to gluon quasi-zero modes, dashed line corresponds to quark quasi-zero modes. The filled circles denote all possible diagrams with normal quark and gluon modes. 
\label{diagrams_effective_mass}}
\end{figure}

If all eigenvalues are sufficiently large to provide Gaussian damping in the functional integral (at small $BR^2$), then formula~\eqref{one-loop_via_det} can be considered as justified. The smaller the eigenvalues, the worse the one-loop approximation becomes, and finally it results in a strong field limit that is upside down . It becomes clear that one has to take into account the mixing between normal and quasi-zero modes, that means going beyond  one-loop approximation of the free energy  at large $BR^2$. Calculation in the effective potential for Abelian self-dual field in the infinite space-time gives a guiding prescription~\cite{Leutwyler:1980ma}, based on the observation that if one evaluates functional integral over normal (nonzero) modes first and accounts for their interaction with zero modes, than the obtained effective action has a finite quadratic in zero modes part. In other words, due to the interactions zero modes gain an effective ``mass'' $\mu_0^2=\bar\varkappa B$,
which provides one with an appropriate Gaussian measure. Here $\bar\varkappa$ is a constant. Schematically,   contribution to effective ``mass'' is shown in Fig.~\ref{diagrams_effective_mass}. The filled circles denote all possible diagrams which include the  propagators of normal gluon  and quark modes. 
For $SU(2)$ gluodynamics the lowest order value of the zero mode "effective mass" takes the value~\cite{Leutwyler:1980ma}
\begin{eqnarray*}
\mu_0^2=\bar\varkappa B,  \ \bar\varkappa=\frac{g^2}{24\pi^2}.
\end{eqnarray*}
Final result for the free energy density in the infinite volume agrees completely with renormalization group estimate~\cite{Pagels:1978dd} and, as it should be, with asymptotic freedom~\cite{Leutwyler:1980ma}.
 
For finite volume case, this observation suggests that, in particular, interaction of normal modes with quasi-zero modes generates a shift
\begin{eqnarray}
\bar\lambda_\mathrm{eff}^2(B,R)=\bar\lambda^2(B,R)+\varkappa(BR^2) B,
\nonumber\\
\label{kappalimit}
\lim_{R\to\infty}\bar\lambda^2(B,R)=0, \ \ \lim_{z\to\infty}\varkappa(z)=\bar\varkappa>0,
\end{eqnarray}
where $\bar\lambda^2(B,R)$ are quasi-zero eigenvalues of operator \eqref{glop}, and the function $\varkappa(BR^2)$ should approach $\bar\varkappa$ both for infinite volume  and in the strong field limit. 

The normal mode propagators, involved into diagrams in Fig.\ref{diagrams_effective_mass},  
can be represented at most as an infinite series over quantum numbers of the modes. Truncation of these series is unreliable  since in general  diagrams are UV-divergent.  Given that,  calculation of the dependence of $\varkappa$ on dimensionless quantity $z=BR^2$  appears to be an extremely complicated task, even in the lowest perturbation order. 
Moreover, in the absence of small expansion parameter,  perturbative expansion can not lead to a decisive result anyway. 

Meanwhile, it seems to be possible to identify the general form of 
$\varkappa(z)$  suitable for qualitative estimate of the available fundamentally different dependencies of the free energy density on the domain size.
Two restrictions for the function $\varkappa(z)$ can be identified.
The first one is given by Eq.~\eqref{kappalimit}. As has already been noted, it follows from the existence of thermodynamic limit and  agreement with the asymptotic freedom and the strong field limit. Another restriction,
 \begin{eqnarray}
\label{kappalimit0}
\lim_{z\to 0}\varkappa(z)=0,
\end{eqnarray}
follows from the scaling of all eigenvalues at small $R$
\begin{eqnarray*}
\lim_{R\to 0}\lambda^2(z)\propto \lim_{R\to 0}1/R^2=0,
\end{eqnarray*}
which means that corrections to the effective action of quasi-zero modes coming from the diagrams (Fig.~\ref{diagrams_effective_mass}) are expected to vanish, in analogy with decoupling of the infinitely heavy particles.

\begin{figure}
\includegraphics[scale=1]{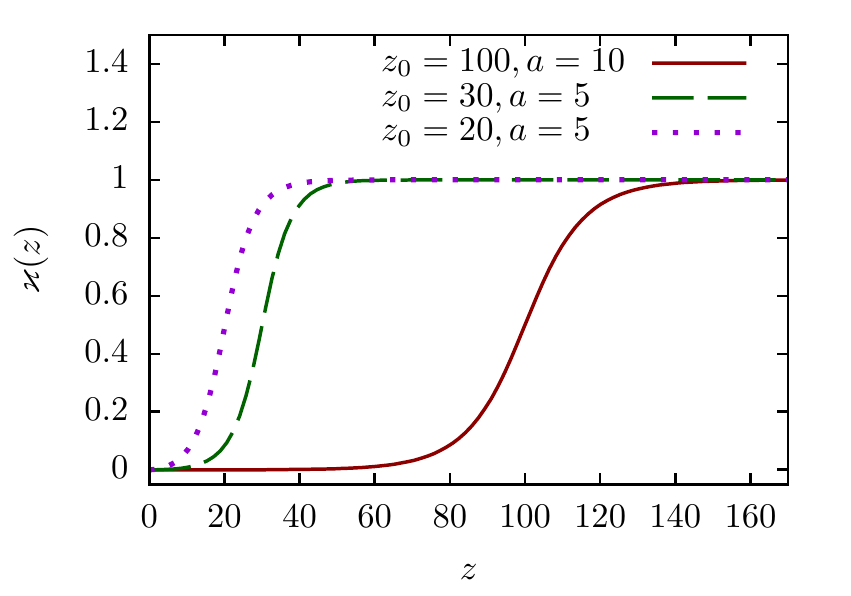}
\caption{The profile of function $\varkappa(z)$ used in Fig.~\ref{gluon_effmass_figure}.\label{kappa_figure}}
\end{figure}

A trial function $\varkappa(z)$   can be taken in the form
\begin{equation}
\label{kappa}
\varkappa(z)=\frac{2}{\pi}\left[\arctan\exp\left(\frac{z-z_0}{a}\right)+\arctan\exp\left(\frac{z_0-z}{a}\right)-2\arctan\exp\left(-\frac{z_0}{a}\right)\right],
\end{equation}
which additionally to above restrictions at $z\to0$ and $z\to\infty$
also reflects change in the behavior of eigenvalues at certain value of $z$, when  all normal eigenvalues start rising as $R$ decreases, see
Fig.~\ref{figure_lambdas}. Function $\varkappa$ is plotted in Fig.~\ref{kappa_figure}.

Incorporation of effective ``mass'' $\varkappa(BR^2)$ 
leads to the following gluon contribution to the effective action
(see Appendix~\ref{appendix_ghost_gluon} for details of calculation):
\begin{eqnarray}
\nonumber
\delta U_\varkappa^\mathrm{gl}(B,R)&=&
4\sum_{k=1}^\infty k \left[\sum_{m=-\frac{k-1}{2}}^\frac{k-1}{2}\Psi^\text{gl}(k,m,0,\frac{\sqrt{3}B}{2},R)-\frac{3}{4}B^2R^4\frac{1}{24}\left(1-\frac{1}{k}-\frac{5}{k^2}\right)\right]\\
\nonumber
&&
-\frac{B^2R^4}{48}(31+30\gamma-30\log 2+30\log R)+\frac{3B^4R^8}{5120}\\
\label{gluon_effpot_mass}
&&+4\sum_{k=0}^\infty \left[(k+1)\log\frac{\lambda_{\uparrow k\frac{k}{2}0}^2(B,R)+\varkappa(BR^2)B}{\lambda_{\uparrow k\frac{k}{2}0}^2(B,R)}+(k+1)\log\frac{\lambda_{\downarrow k\frac{-k}{2}0}^2(B,R)-\varkappa(BR^2)B}{\lambda_{\downarrow k\frac{-k}{2}0}^2(B,R)}\right].
\end{eqnarray}

\begin{figure}
\includegraphics[scale=1]{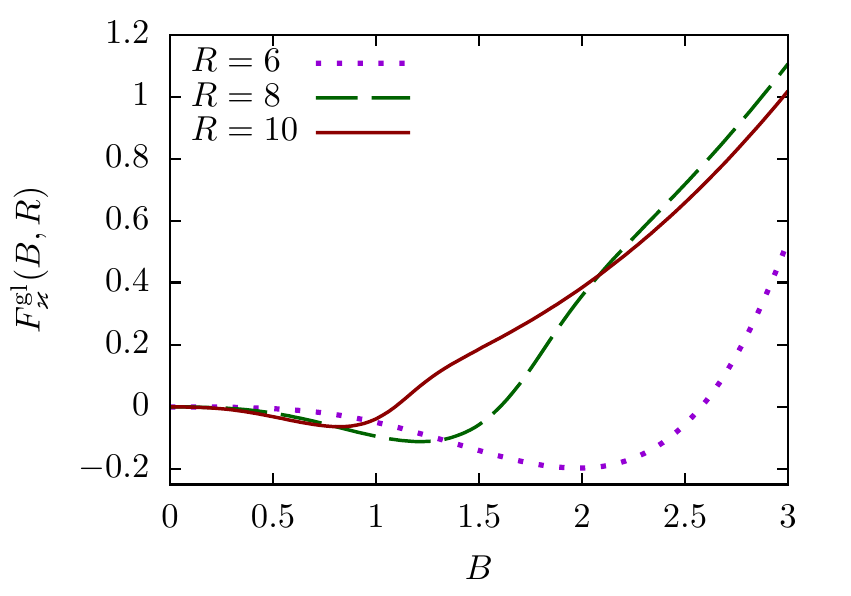}\hfill
\includegraphics[scale=1]{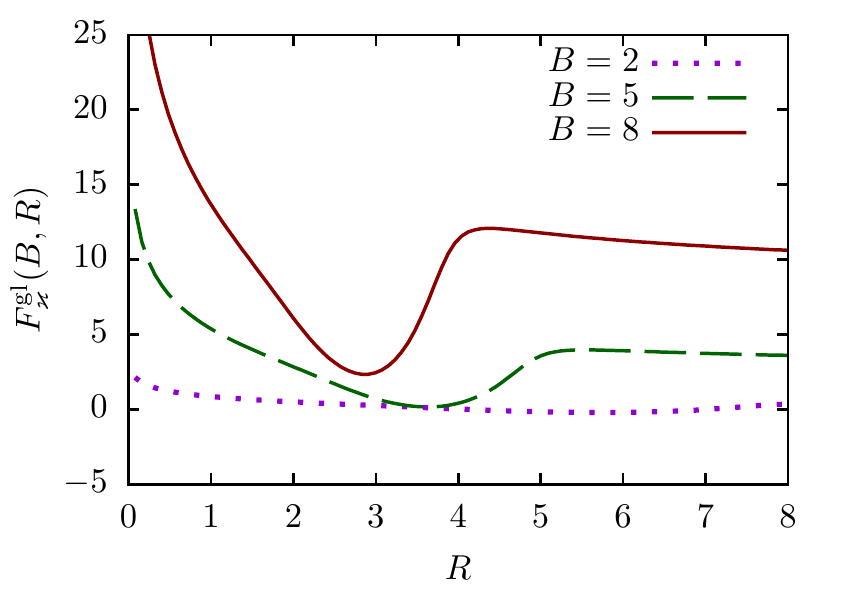}
\caption{Improved gluon contribution to the free energy with effective "mass" $\varkappa$ given by Eq.~\eqref{kappa}. Parameters of $\varkappa$ are taken such that the minimum would be well-pronounced,$z_0=100$ and $a=10$. All quantities are measured in units of renormalization scale $\mu$, Eq.~\eqref{muscale}.
\label{gluon_effmass_figure}}
\end{figure}

The corresponding free energy density is plotted in Fig.~\ref{gluon_effmass_figure}. The correct behavior of the free energy density for $BR^2\to\infty$,
 consistent with asymptotic freedom and existence of thermodynamic limit,
 is restored.  It is seen that free energy density acquires a minimum at intermediate values of field strength and domain size. 

Free energy density for pure $SU(3)$ gluodynamics  for finite domain of Abelian (anti-)self-dual gluon fields  is given by sum of Eqs.~\ref{ghosts_effpot} and~\ref{gluon_effpot_mass}. The result is illustrated in left-hand side of Fig.~\ref{effpot_total_figure}. 
Existence of the minimum of the free energy density as a function of two variables is clearly seen.

\subsection{Quark contribution}

To complete the calculation for full QCD with massless quarks, we have to study the quark contribution to the  effective potential 
\begin{equation*}
\delta U^\mathrm{q}(B,R)=-N_f\mathrm{Tr}\ln \frac{i \hspace*{-0.3em}\not \hspace*{-0.3em}D}{\left. i \hspace*{-0.3em}\not \hspace*{-0.3em}D\right|_{B=0}}=-N_f\sum_{k,j,n}\mathrm{Tr}\ln\frac{i\lambda_{kjn}(B,R)}{i\lambda_{kjn}(0,R)}=N_f\left.\frac{d}{ds}\zeta^\mathrm{q}(s)\right|_{s=0},
\end{equation*}
where $N_f$ is number of quark flavors, $\lambda(B,R)$ are eigenvalues of Dirac operator in a spherical domain of radius $R$ with homogeneous Abelian (anti-)self-dual field with bag boundary condition (see Appendix~\ref{appendix_quark}).

\begin{figure}
\includegraphics[width=0.49\textwidth]{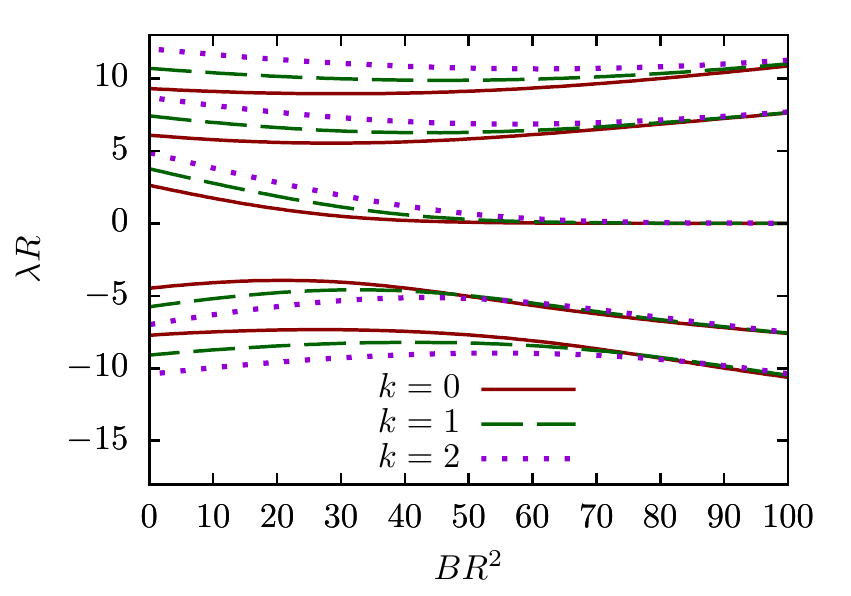}
\caption{Dependence of quark eigenvalues on $BR^2$. At $BR^2\to \infty$, infinite number of zero modes emerges.  Normal modes come in positive-negative pairs $\lambda^{\pm}$. Note that $\lambda^{+}\slashed{=}-\lambda^{-}$ for finite $BR^2$, but the equality is restored in the limit $BR^2\to\infty$. An infinite number of quasi-zero modes are not chiral for finite $R$ and  become chiral zero-modes for asymptotically large $BR^2$.  \label{figure_lambdasq}}
\end{figure}

Zeta function $\zeta^\mathrm{q}(s)$ can be split into two parts~\cite{Deser:1997nv} 
\begin{eqnarray}
\label{zeta_split}
&&\zeta^\mathrm{q}(s)=\cos(\pi s)\zeta_{\slashed{D}^2}\left(\frac{s}{2}\right)-i\sin(\pi s)\eta(s),\\
\nonumber
&&\left.\frac{d}{ds}\zeta^\mathrm{q}(s)\right|_{s=0}= \frac12 \zeta_{\slashed{D}^2}'(0)- i\pi \eta(0).
\end{eqnarray}
For the purpose of the present study we need only $\zeta_{\slashed{D}^2}$. Parity-odd term $\eta(s)$ contributes to the imaginary part of the effective potential and can be related to the $U_{\rm A}(1)$ anomalous breakdown~\cite{Kalloniatis:2003sa}. 
Parity-even part $\zeta_{\slashed{D}^2}$ can be written in the following form:
\begin{eqnarray}
\label{zeta_integral_rep}
&&\zeta_{\slashed{D}^2}(s)=\mathrm{Tr}\frac{\sin\pi s}{\pi} \sum_{k=0}^\infty \sum_{j_3=-\frac{k+1}{2}}^{\frac{k+1}{2}} (k+1) \int_0^\infty \frac{dt}{t^{2s}}\frac{d}{dt}\Psi^\mathrm{q}(k+1,j_3,t,\hat{B},R),\\
\nonumber
&&\Psi^\mathrm{q}(k+1,j,t,\hat{B},R)=\log \frac{A(-it,k,j_3,\hat{B},R)A(it,k,j_3,\hat{B},R)}{A(-it,k,j_3,0,R)A(it,k,j_3,0,R)},
\end{eqnarray}
where $A(\lambda,k,j,B,R)=0$ is the equation for eigenvalues.
Proceeding in the same manner as in the previous section (see Appendix~\ref{appendix_quark_zeta} for details), we arrive at the expression
\begin{eqnarray}
\nonumber
\delta U^\mathrm{q}(B,R)&=&
N_f \left\{-\frac{1}{2}\mathrm{Tr}\sum_{k=1}^\infty (k+1) \left[\sum_{j=-\frac{k-1}{2}}^\frac{k-1}{2}\Psi^\text{q}(k,j,0,\hat{B},R)-\hat{B}^2R^4\frac{1}{12}\left(1-\frac{2}{k^2}\right)\right]\right.\\
\label{quarks_effpot}
&&+\left.\frac{B^2R^4}{144}(5+6\gamma-6\log 2+\pi^2+6\log R)-
B^4R^8\frac{1}{30720}\vphantom{\sum_\frac11^\frac11}\right\}.
\end{eqnarray}
\begin{figure}
\includegraphics[scale=1]{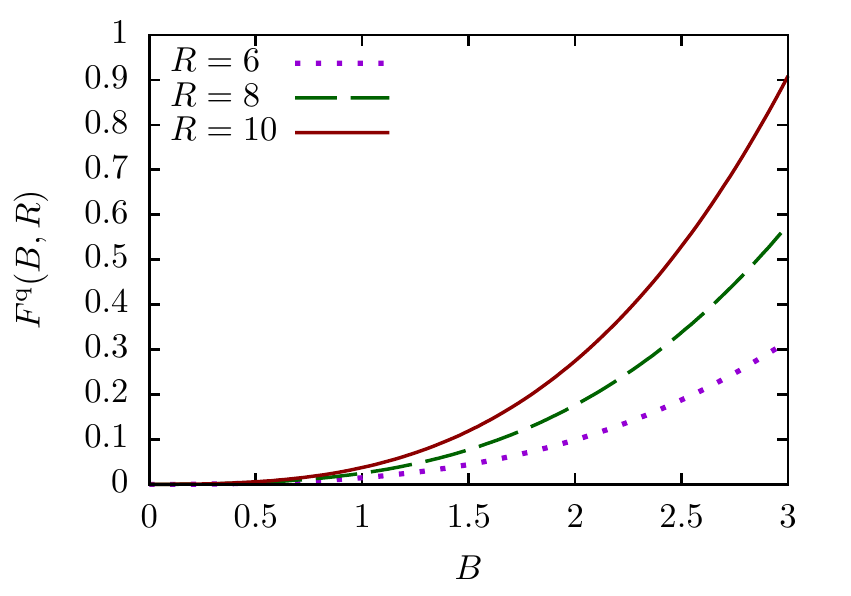}\hfill
\includegraphics[scale=1]{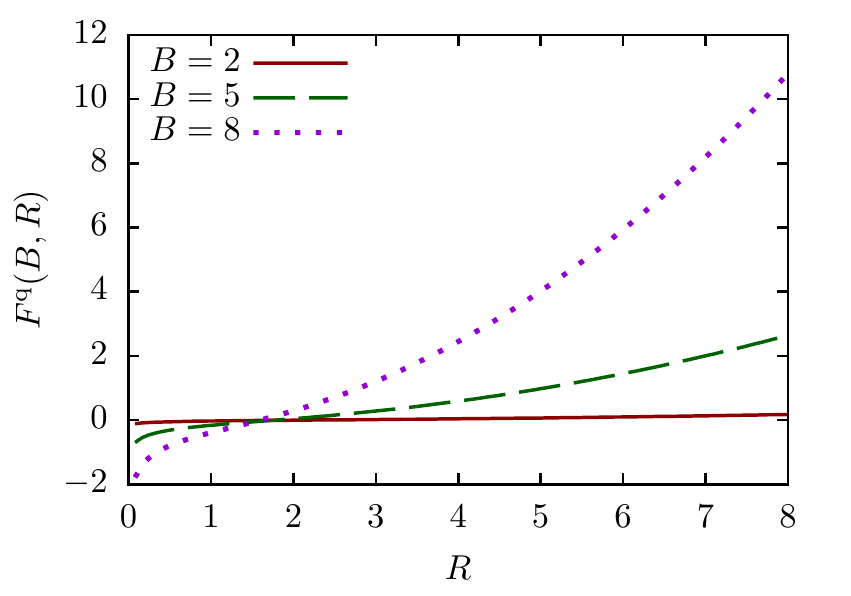}
\caption{Dependence of one-loop quark contribution to the free energy density  on $R$ and $B$, $N_f=1$. All quantities are given in units of renormalization scale $\mu$, Eq.~\eqref{muscale}. The strong-field and large-size regimes are incorrect. \label{quarks_figure}}
\end{figure}
The quark contribution to the free energy density given by Eq.~\eqref{quarks_effpot} and shown in Fig.~\ref{quarks_figure} exhibits, due to the quark quasi-zero modes, the inconsistency with the correct strong field and thermodynamic limits similar to inconsistencies in the  gluon effective potential given by \eqref{gluons_effpot} and illustrated in Fig.~\ref{gluon_figure}.

\begin{figure}
\includegraphics[scale=1]{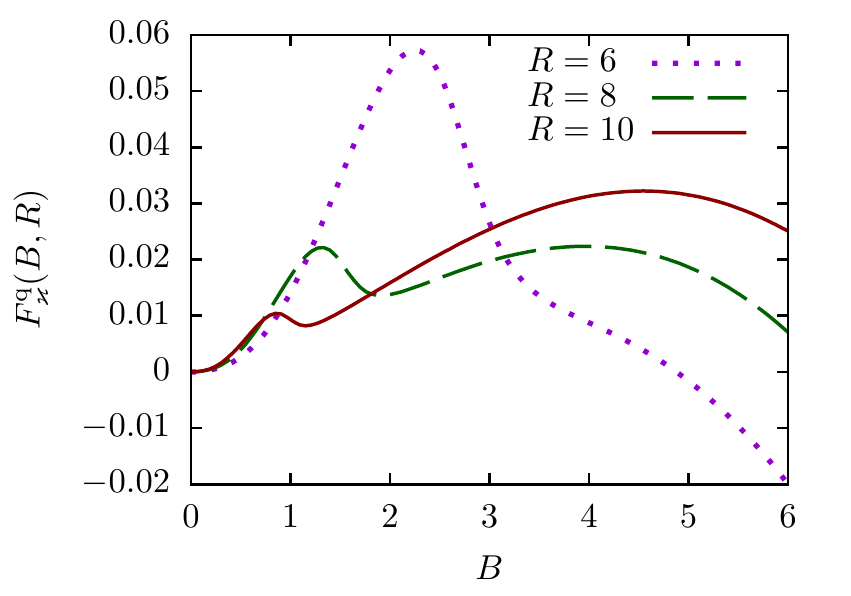}\hfill
\includegraphics[scale=1]{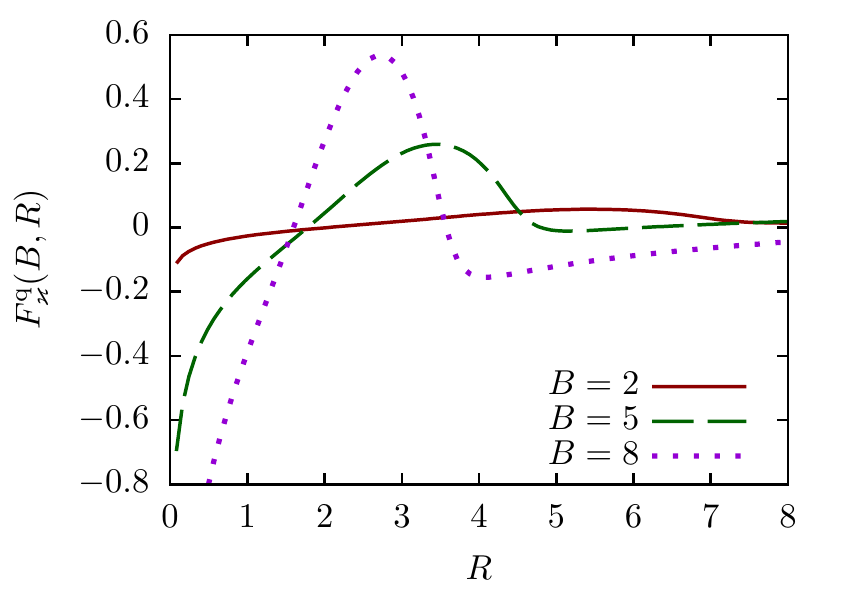}
\caption{Improved quark contribution to the free energy density $R$ and $B$, $N_f=1$, in units of renormalization scale $\mu$, Eq.~\eqref{muscale}. Function $\varkappa$ is given by Eq.~\eqref{kappa} ($z_0=100,a=10$).   correct strong-field and large-size limits are restored. 
\label{quarks_effmass_figure}}
\end{figure}

Following the reasoning analogous to the case of gluons in the previous subsection, the improved calculation of the  quark contribution has to take into account generation of the  effective ``mass'' for quasi-zero eigenmodes of quarks due to the interaction of quark quasi-zero modes with normal gluon and normal quark modes (see right-hand side diagram in 
Fig.~\ref{diagrams_effective_mass}). The improved quark contribution  reads
\begin{eqnarray}
\nonumber
\delta U_\varkappa^\mathrm{q}(B,R)&=&
N_f \left\{-\frac{1}{2}\mathrm{Tr}\sum_{k=1}^\infty (k+1) \left[\sum_{j_3=-\frac{k-1}{2}}^\frac{k-1}{2}\Psi^\text{q}(k,j,0,\hat{B},R)-\hat{B}^2R^4\frac{1}{12}\left(1-\frac{2}{k^2}\right)\right]\right.\\
\nonumber
&&\left.+\frac{B^2R^4}{144}(5+6\gamma-6\log 2+\pi^2+6\log R)-
B^4R^8\frac{1}{30720}\vphantom{\sum_\frac11^\frac11}\right\}\\
\label{quarks_effpot_mass}
&&-\frac{N_f}{2}\mathrm{Tr}\sum_{k=0}^\infty \left[(k+1)\log\frac{\lambda_{k\frac{k}{2}0}^2(B,R)+\varkappa(BR^2)B}{\lambda_{k\frac{k}{2}0}^2(B,R)}+(k+1)\log\frac{\lambda_{k\frac{-k}{2}0}^2(B,R)-\varkappa(BR^2)B}{\lambda_{k\frac{-k}{2}0}^2(B,R)}\right],
\end{eqnarray}
similarly to Eq.~\eqref{gluon_effpot_mass}.
The effective potential is shown in Fig.~\ref{quarks_effmass_figure}.

\begin{figure}
\includegraphics[scale=1]{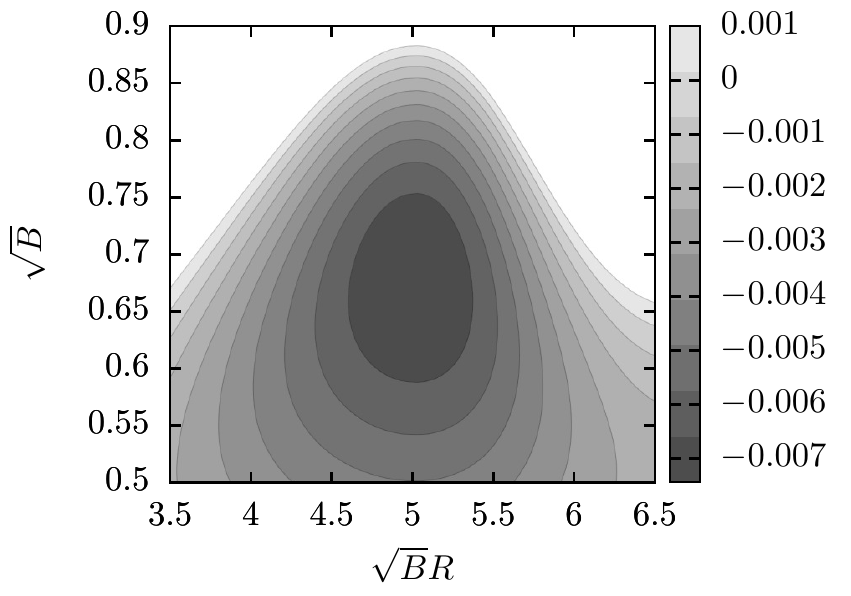}
\includegraphics[scale=1]{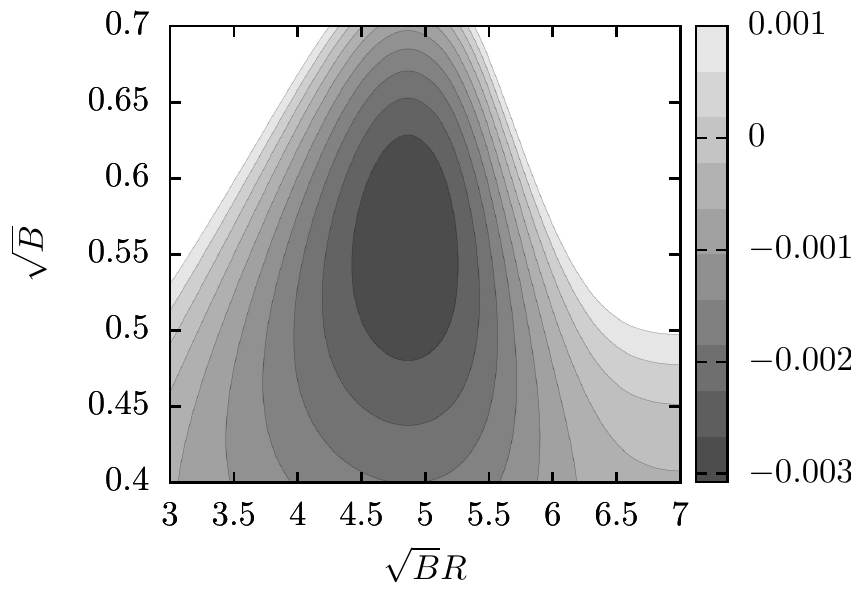}
\caption{Total quantum correction to the free energy density $F_\varkappa(B,R)$ given by Eq.~\eqref{effpot_total_kappa}  with $\varkappa$ given by Eq.~\eqref{kappa} ($z_0=30,a=5$). Left-hand side plot is for  pure gluodynamics ($N_f=0$), right-hand side plot is for full QCD with two massless quark flavors ($N_f=2$), in units of renormalization scale 
$\mu$.
\label{effpot_total_figure}}
\end{figure}

Combining ghost, improved gluon and quark contributions, one finds
\begin{eqnarray}
\nonumber
&&\delta U_\varkappa(B,R)=\delta U_\varkappa^\mathrm{gl}(B,R)+\delta U^\mathrm{gh}(B,R)+\delta U_\varkappa^\mathrm{q}(B,R)=\\
\nonumber
&&=4\sum_{k=1}^\infty k \left[\sum_{m=-\frac{k-1}{2}}^\frac{k-1}{2}\Psi^\text{gl}(k,m,0,\frac{\sqrt{3}B}{2},R)-\frac{3}{4}B^2R^4\frac{1}{24}\left(1-\frac{1}{k}-\frac{5}{k^2}\right)\right]\\
\nonumber
&&+4\sum_{k=0}^\infty \left[(k+1)\log\frac{\lambda_{\mathrm{gl},k\frac{k}{2}0}^2(B,R)+\varkappa(BR^2)B}{\lambda_{\mathrm{gl},k\frac{k}{2}0}^2(B,R)}+(k+1)\log\frac{\lambda_{\mathrm{gl},k\frac{-k}{2}0}^2(B,R)-\varkappa(BR^2)B}{\lambda_{\mathrm{gl},k\frac{-k}{2}0}^2(B,R)}\right]\\
\nonumber
&&-4\sum_{k=1}^\infty k \left[\sum_{m=-\frac{k-1}{2}}^\frac{k-1}{2}\Psi^\text{gh}(k,m,0,\frac{\sqrt{3}B}{2},R)-\frac{3}{4}B^2R^4\frac{1}{48}\left(1-\frac{1}{k}+\frac{1}{k^2}\right)\right]\\
\nonumber
&&-\frac{N_f }{2}\mathrm{Tr}\sum_{k=1}^\infty (k+1) \left[\sum_{j_3=-\frac{k-1}{2}}^\frac{k-1}{2}\Psi^\text{q}(k,j,0,\hat{B},R)-\hat{B}^2R^4\frac{1}{12}\left(1-\frac{2}{k^2}\right)\right]\\
\nonumber
&&-\frac{N_f}{2}\mathrm{Tr}\sum_{k=0}^\infty \left[(k+1)\log\frac{\lambda_{\mathrm{q},k\frac{k}{2}0}^2(B,R)+\varkappa(BR^2)B}{\lambda_{\mathrm{q},k\frac{k}{2}0}^2(B,R)}+(k+1)\log\frac{\lambda_{\mathrm{q},k\frac{-k}{2}0}^2(B,R)-\varkappa(BR^2)B}{\lambda_{\mathrm{q},k\frac{-k}{2}0}^2(B,R)}\right]\\
\label{effpot_total_kappa}
&&-\frac{B^2R^4}{48}(29+33\gamma-33\log 2+33\log R)+\frac{3B^4R^8}{10120}+N_f \left\{\frac{B^2R^4}{144}(5+6\gamma-6\log 2+\pi^2+6\log R)-
\frac{B^4R^8}{30720}\vphantom{\sum_\frac11^\frac11}\right\}.
\end{eqnarray}
Corresponding free energy density is shown in right-hand side of Fig.~\ref{effpot_total_figure}.   The total free energy density   demonstrates a well-pronounced minimum 
as a function of field strength and domain size. Quark contribution does not change the result of pure gluodynamics qualitatively, though the field strength at the minimum is considerably reduced.

\subsection{One-loop beta function}

The effective action $U=U^{\rm cl}+\delta U_{\varkappa}$ should not depend on renormalization  scale  $\mu$~\cite{Wiesendanger:1993mw,Cognola:1993xy}, that is 
\begin{equation}
\mu \frac{d}{d\mu}U=0,
\label{RGg}
\end{equation}
with  classical action ($V_R$ is four-dimensional ball of radius $R$)
\begin{equation*}
U^\textrm{cl}=\int_{V_R} d^4x \frac{1}{4g^2}\left(B^a_{\mu\nu}\right)^2=\frac{\pi^2B^2R^4}{2g^2},
\end{equation*}
and $\delta U_{\varkappa}$ given by Eq.\eqref{effpot_total_kappa}. 
Since only terms containing $\log R$ contribute to the Eq.\eqref{RGg}, one obtains
\begin{equation*}
\mu\frac{d}{d\mu}\left[\frac{\pi^2B^2R^4}{2g^2}+\frac{B^2R^4}{16}\left(-11+\frac{2}{3}N_f\right)\log R\right]=0,
\end{equation*}
that is equivalent to the equation
\begin{equation*}
\mu\frac{d}{d\mu}g=-\frac{g^3}{16\pi^2}\left(11-\frac{2}{3}N_f\right),
\end{equation*}
exposing the correct one-loop $\beta$-function of QCD with $N_{\rm f}$
quark flavors.

\section{Discussion}

We have studied, as far as it has been possible with analytical methods, an influence  of the finite size effects on the vacuum free energy density of full QCD with $N_{\rm f}$ massless flavors in the presence of
homogeneous  (anti-)self-dual Abelian background gluon field.
The most essential result is illustrated in the right-hand side 
Fig.\ref{effpot_total_figure}, where the zero temperature free energy density of the four-dimensional spherical domain  is plotted as a function of the background field strength $B$ and domain radius $R$. 
It indicates that the quantum correction to the free energy density  may have a minimum at finite values of $B$ and $R$.  
In the domain wall network representation of the vacuum mean field,  existence of this minimum means that in the statistically dominant networks the individual domains should have finite size varying near the mean value,  and infinite growth of individual domains  is prohibited by the minimization of the overall free energy of the configuration.
This suggests that the domain wall ensemble should include an infinite number of configurations with degenerate energy. 

The character of this energy-driven disorder strongly depends on the details of behavior of the free energy density in the vicinity of the minimum.  One may expect that shallow and flat profile near the minimum may lead to strong variations of the geometrical shape as well as deviations of the field strength from the mean value, which will allow the presence of highly irregular networks among the dominant configurations, characterized by the strong entanglement of domains, etc. 
The  deep and steep  profile would assume that spatially periodic networks should dominate, bringing the long-range (periodic) order into the mean field configurations.

The result of the present paper, Fig.\ref{effpot_total_figure}, is certainly has a status of  preliminary rough estimate, since the validity of one-loop approximation is indeterminate, especially due to the indefiniteness of the treatment of interaction between quasi-zero and normal modes.     
A straightforward numerical calculation, within the lattice approximation for instance, could be useful.  A difficulty for lattice calculation  can be caused by the  non-standard boundary conditions, which have to represent the physical boundary of a domain.  However, this possibility  does not look like hopeless in view of the recent lattice QCD calculations for rotating strongly interacting matter, where Dirichlet and Neumann boundary conditions have been implemented~\cite{Braguta:2020eis,Yamamoto:2013zwa}.

\acknowledgments

We acknowledge useful discussions with Michael Bordag, Irina Pirozhenko,
Victor Braguta and Artem Roenko.

\appendix

\section{Zeta function for ghost and gluon fields\label{appendix_ghost_gluon}}
\subsection{Ghosts}
We start with the expression for $\zeta^\text{gh}(s)$ given by~\eqref{ghosts_zeta_start}. Suitable asymptotic expansion for hypergeometric function $M$ at large $k$ and fixed $j/k$ can be found with the help of the method described in Ref.~\cite{Olver:1997}, Chapter 10, \S 9. We obtain
\begin{eqnarray}
\nonumber
&&M\left(\frac{m+1}{2}+\frac{k}{2}-j+n+\frac{k^2t^2}{4},1+k+m,z\right)\sim\\
\label{Kummer_asymptotic_expansion}
&&\frac{2^{k+m}\Gamma(k+m+1)}{k^{k+m}\sqrt{2\pi k}(1+t^2z)^{\frac{1}{4}}}\exp\left(k\sqrt{1+t^2z}-(k+m) \log\left[\sqrt{1+t^2z}+1\right]-\frac{2j}{k}\frac{z}{\sqrt{1+t^2z}+1}+\frac{z}{2}\right)\sum_{i=0}^\infty \frac{A_i(z)}{k^i}
\end{eqnarray}
for fixed $z\geqslant 0$ and arbitrary constants $m,n$.
The coefficients $A_i(z)$ are found with the help of recursion relation ($i\geqslant 0, A_0=1$)
\begin{equation*}
A_{i+1}(z)=-\frac{1}{2}\phi-\frac{1}{2}\frac{dA_i}{dz}\frac{dz}{d\xi}+\int dz \frac{d\xi}{dz} \left(\psi +\frac{1}{2}\frac{d\phi}{dz}\frac{dz}{d\xi}-\frac{1}{4}\phi^2\right)A_i,
\end{equation*}
where
\begin{equation*}
\frac{d\xi}{dz}=\frac{1}{2z}\sqrt{1+t^2z},\quad \phi=\frac{2}{1+t^2z}\left(m-\frac{2j}{k}\right),\quad \psi=\frac{m^2+4nz}{1+t^2z}+\frac{4t^2z+4z^2-t^4z^2+8t^2z^3+4t^4z^4}{(1+t^2z)^3}.
\end{equation*}
The constants of integration are fixed by the requirement
\begin{equation*}
\lim_{t\to\infty} A_{i}(z)=0,\quad i\geqslant 1.
\end{equation*}

Asymptotic expansion for modified Bessel function $I$ is given by (see~\cite{NIST:DLMF}, 10.41.3, 10.41.7, 10.41.9)
\begin{equation}
\label{Bessel_asymptotic_expansion}
I_k(kt)=\frac{1}{\sqrt{2\pi k}(1+t^2)^{\frac{1}{4}}}\exp\left(k\sqrt{1+t^2}+k\log\frac{t}{1+\sqrt{1+t^2}}\right)\sum_{i=0}^\infty \frac{U_i(p)}{k^i},
\end{equation}
where
\begin{equation*}
U_{k+1}(p)=\frac12 p^2(1-p^2)U_k'(p)+\frac18 \int_0^p dp'(1-5p'^2)U_k(p'),\quad p=\sqrt{1+t^2}.
\end{equation*}

Now, we substitute formulas~\eqref{Kummer_asymptotic_expansion} and~\eqref{Bessel_asymptotic_expansion} into~$\Psi^\mathrm{gh}$ in formula~\eqref{ghosts_zeta_start}, expand it in powers of $k^{-1}$, sum over $m$ and compute derivative with respect to $t$. We arrive at the following expressions for $u_i^\mathrm{gh}$
\begin{eqnarray*}
u_0^\mathrm{gh}(t,B,R)&=&B^2R^4\frac{R^4t^4-2R^2t^2-4+4\left(1+R^2t^2\right)^{1/2}}{24R^4 t^4\left(1+R^2t^2\right)^{1/2}},\\
u_1^\mathrm{gh}(t,B,R)&=&-B^2R^4\frac{1+3R^2t^2}{48\left(1+R^2t^2\right)},\\
u_2^\mathrm{gh}(t,B,R)&=&\frac{B^2R^4}{192\left(1+R^2t^2\right)^{1/2}}\left[\frac{16}{R^2t^2}+\frac{R^2t^2(32+7R^2t^2)}{(1+R^2t^2)^3}+32\frac{1-\left(1+R^2t^2\right)^{1/2}}{R^4t^4}\right]\\
&&+\frac{B^4R^8}{1920\left(1+R^2t^2\right)^{1/2}}
\left[-\frac{128}{R^6t^6}+\frac{32}{R^4t^4}-\frac{16}{R^2t^2}+\frac{10+13R^2t^2}{(1+R^2t^2)^2}-256\frac{1-\left(1+R^2t^2\right)^{1/2}}{R^8t^8}\right].
\end{eqnarray*}
The sums over $k$ are calculated for $\mathfrak{R}s>1$ and analytically continued in terms of Riemann $\zeta$ function
\begin{equation*}
\sum_{k=1}^\infty \frac{k^{1-2s}}{k^i}=\zeta(i-1+2s).
\end{equation*}
to the strip $0<\mathfrak{R}s<1$ where integrals over $t$ converge at $t\to 0$.
One obtains
\begin{eqnarray*}
\sum_{k=1}^\infty k^{1-2s} \int_0^\infty \frac{dt}{t^{2s}}\frac{d}{dt}\sum_{i=0}^2 \frac{u_i^{\mathrm{gh}}(t,B,R)}{k^i}&=&
B^2R^{4+2s}\left\{\vphantom{\frac11}\right.\zeta(-1+2s)\frac{-\Gamma(2-s)\Gamma(1/2+s)}{24\sqrt{\pi}(2+s)}+
\zeta(2s)\frac{\pi s(1-2s)}{48\sin\pi s}\\
&&\left.+\zeta(1+2s)\frac{\left(6-s(2+s)(11+5s)\right)\Gamma(1-s)\Gamma(3/2+s)}{72\sqrt{\pi}(2+s)}
\right\}\\
&&-B^4R^{8+2s}\zeta(1+2s)\frac{s\Gamma(2-s)\Gamma(3/2+s)}{480\sqrt{\pi}(4+s)}.
\end{eqnarray*}
The expansion of counterterms in powers of $s$ around $s=0$ is
\begin{eqnarray*}
\frac{\sin\pi s}{\pi}\sum_{k=1}^\infty k^{1-2s} \int_0^\infty \frac{dt}{t^{2s}}\frac{d}{dt}\sum_{i=0}^2 \frac{u_i^{\mathrm{gh}}(t,B,R)}{k^i}&=&B^2R^4\left[-\frac{1}{96}+\frac{1}{288}(4-6\gamma+\log 64-6\log R)s\right]+
B^4R^8\left[-\frac{1}{7680}s\right]\\
&&+O(s^2),
\end{eqnarray*}
where $\gamma=0.5772156649\dots$ is Euler's constant.
For $su(3)$ and $\breve{n}$ given by~\eqref{n_specific}
\begin{equation*}
\mathrm{Tr}\breve{n}^2=3,\quad \mathrm{Tr}\breve{n}^4=4\frac{9}{16}=\frac{9}{4}.
\end{equation*}
Since the whole spectrum is invariant with respect to $B\to -B$, trace over color leads to factor 4. Evaluating the derivative of $\zeta^\mathrm{gh}(s)$ with respect to $s$ at $s=0$ one arrives at the expression~\eqref{ghosts_effpot}.

\subsection{Gluons}

With Dirichlet boundary condition for color-charged modes the whole set of eigenvalues is determined by the equations~\cite{Kalloniatis:2001dw}
\begin{equation*}
M\left(\frac{k}{2}+1-m\pm 1-\frac{\lambda^2}{2n_a B},k+2,\frac{n_a BR^2}{2}\right)=0,\quad k=0,1,2,\dots,\quad m=-\frac{k}{2},-\frac{k}{2}+1,\dots,\frac{k}{2},
\end{equation*}
and every solution of these equations with given $k,m$ and $a$ is $2(k+1)$-degenerate. Repeating the procedure carried out for ghosts, one finds
\begin{eqnarray}
\label{zeta_gl}
\nonumber
\zeta^\text{gl}(s)&=&\mathrm{Tr}\frac{\sin \pi s}{\pi}\sum_{k=0}^\infty \sum_{m=-\frac{k}{2}}^\frac{k}{2}2(k+1) \int_0^\infty \frac{dt}{t^{2s}}\frac{d}{dt}\Psi^\mathrm{gl}(k+1,m,t,\breve{B},R)\\
&=&2\mathrm{Tr}\left\{\frac{\sin\pi s}{\pi}\sum_{k=1}^\infty k^{1-2s} \int_0^\infty \frac{dt}{t^{2s}}\frac{d}{dt}\left[\sum_{m=-\frac{k-1}{2}}^\frac{k-1}{2}\Psi^\mathrm{gl}(k,m,kt ,\breve{B},R)-\sum_{i=0}^2  \frac{u_i^{\mathrm{gl}}(t,\breve{B},R)}{k^i}\right]\right.\\
\nonumber
&&\left.
+\frac{\sin\pi s}{\pi}\sum_{k=1}^\infty k^{1-2s} \int_0^\infty \frac{dt}{t^{2s}}\frac{d}{dt}\sum_{i=0}^2 \frac{u_i^{\mathrm{gl}}(t,\breve{B},R)}{k^i}
\right\},
\\
\nonumber
\Psi^\mathrm{gl}(k,m,t,B,R)&=&\log\frac{\exp\left(-\frac{BR^2}{2}\right)M\left(\frac{k+1}{2}-m+1+\frac{t^2}{2B},k+1,\frac{BR^2}{2}\right)M\left(\frac{k+1}{2}-m-1+\frac{t^2}{2B},k+1,\frac{BR^2}{2}\right)}{\left(k!\left(\frac{t R}{2}\right)^{-k}I_{k}(t R)\right)^2}.
\end{eqnarray}
We substitute formulas~\eqref{Kummer_asymptotic_expansion} and~\eqref{Bessel_asymptotic_expansion} into~$\Psi^\mathrm{gl}$, expand it in powers of $k^{-1}$, sum over $m$ and compute derivative with respect to $t$.
Coefficients of asymptotic expansion $u_i^\mathrm{gl}$ are given by
\begin{eqnarray*}
u_0^\mathrm{gl}(t,B,R)&=&B^2R^4\frac{R^4t^4-2R^2t^2-4+4\left(1+R^2t^2\right)^{1/2}}{12R^4 t^4\left(1+R^2t^2\right)^{1/2}},\\
u_1^\mathrm{gl}(t,B,R)&=&-B^2R^4\frac{1+3R^2t^2}{24\left(1+R^2t^2\right)},\\
u_2^\mathrm{gl}(t,B,R)&=&\frac{B^2R^4}{96\left(1+R^2t^2\right)^{1/2}}\left[-\frac{80}{R^2t^2}+\frac{R^2t^2(32+7R^2t^2)}{(1+R^2t^2)^3}-160\frac{1-\left(1+R^2t^2\right)^{1/2}}{R^4t^4}\right]\\
&&+\frac{B^4R^8}{960\left(1+R^2t^2\right)^{1/2}}
\left[-\frac{128}{R^6t^6}+\frac{32}{R^4t^4}-\frac{16}{R^2t^2}+\frac{10+13R^2t^2}{(1+R^2t^2)^2}-256\frac{1-\left(1+R^2t^2\right)^{1/2}}{R^8t^8}\right].
\end{eqnarray*}
One obtains
\begin{eqnarray*}
\sum_{k=1}^\infty k^{1-2s} \int_0^\infty \frac{dt}{t^{2s}}\frac{d}{dt}\sum_{i=0}^2 \frac{u_i^{\mathrm{gl}}(t,B,R)}{k^i}&=&B^2R^{4+2s}\left\{\vphantom{\frac11}\right.\zeta(-1+2s)\frac{-\Gamma(2-s)\Gamma(1/2+s)}{12\sqrt{\pi}(2+s)}+
\zeta(2s)\frac{\pi s(1-2s)}{24\sin\pi s}\\
&&+
\left.
\zeta(1+2s)\frac{\left(30+s(2+s)(11+5s)\right)\Gamma(1-s)\Gamma(3/2+s)}{36\sqrt{\pi}(2+s)}
\right\}\\
&&-B^4R^{8+2s}\zeta(1+2s)\frac{s\Gamma(2-s)\Gamma(3/2+s)}{240\sqrt{\pi}(4+s)}.
\end{eqnarray*}
Now, counterterms can be expanded in powers of $s$
\begin{eqnarray*}
\frac{\sin\pi s}{\pi}\sum_{k=1}^\infty k^{1-2s} \int_0^\infty \frac{dt}{t^{2s}}\frac{d}{dt}\sum_{i=0}^2 \frac{u_i^{\mathrm{gl}}(t,B,R)}{k^i}&=&
B^2R^4\left[\frac{5}{48}+\frac{1}{144}(31+30\gamma-30\log 2+30\log R)s\right]+
B^4R^8\left[-\frac{1}{3840}s\right]\\
&&+O(s^2).
\end{eqnarray*}
Finally,
\begin{eqnarray*}
\delta U^\mathrm{gl}(B,R)=-\frac{1}{2}\left.\frac{d}{ds}\zeta^\text{gl}(s)\right|_{s=0}&=&4\sum_{k=1}^\infty k \left[\sum_{m=-\frac{k-1}{2}}^\frac{k-1}{2}\Psi^\text{gl}(k,m,0,\frac{\sqrt{3}B}{2},R)-\frac{3}{4}B^2R^4\frac{1}{24}\left(1-\frac{1}{k}-\frac{5}{k^2}\right)\right]\\
&&-\frac{B^2R^4}{48}(31+30\gamma-30\log 2+30\log R)+\frac{3B^4R^8}{5120}.
\end{eqnarray*}

\subsection{Contribution of quasi-zero modes}
In this section, we calculate the contribution of gluon quasi-zero modes $\lambda_{\uparrow k\frac{k}{2}0}$ to the effective potential. These modes correspond to the smallest-magnitude solutions to the equations
\begin{equation}
\label{gluon_quasizero_modes_equation}
M\left(-\frac{\lambda^2}{2|\breve{B}|},k+2,\frac{|\breve{B}|R^2}{2}\right)=0,\quad
k=0,1,2,\dots
\end{equation}
that reduce to
\begin{equation*}
\left(\frac{\lambda R}{2}\right)^{-k-1}J_{k+1}(\lambda R)=0,\quad k=0,1,\dots
\end{equation*}
at $B\to 0$.
The contribution of gluon quasi-zero modes to the effective potential can be expressed as
\begin{eqnarray}
\nonumber
&&\delta U^{\textrm{gl}(0)}(B,R)=\frac{1}{2}\textrm{Tr} \log\frac{\lambda_{\uparrow k\frac{k}{2}0}^2(B,R)}{\lambda_{k0}^2(0,R)}=\left.-\frac{1}{2}\frac{d}{ds}\zeta^{\textrm{gl}(0)}(s)\right|_{s=0},\\
\label{zeta_quasi-zero_gl}
&&\zeta^{\textrm{gl}(0)}(s)=2\mathrm{Tr}\sum_{k=0}^\infty (k+1) \left( \lambda_{\uparrow k\frac{k}{2}0}^{-2s}(B,R)-\lambda_{k0}^{-2s}(0,R) \right),
\end{eqnarray}
where factor $2$ in the definition of $\zeta^{\textrm{gl}(0)}(s)$ originates from two polarizations of quasi-zero gluon modes, and color trace yields factor four. For the sake of brevity we omit color eigenvalue and restore it in the final answer ($B\to \frac{\sqrt{3}}{2}B$ for adjoint representation of $su(3)$)

To continue $\zeta^{\textrm{gl}(0)}(s)$ to $s\to 0$, we add and subtract several terms of asymptotic expansion in $k$ found with the help of formula~\eqref{zero_asymptotic_expansion}
\begin{eqnarray*}
\zeta^{\textrm{gl}(0)}(s)&=&8\sum_{k=0}^\infty \left[(k+1) \left( \lambda_{\uparrow k\frac{k}{2}0}^{-2s}(B,R)-\lambda_{k0}^{-2s}(0,R) \right)\right.\\
&&\left.-\left(k+1\right)^{-2s}R^{2s}\left(BR^2s-2\alpha BR^2s(1+s)(k+1)^{-2/3}+BR^2\frac{s}{4}\left(4+BR^2(1+2s)\right)(k+1)^{-1}\right)\right]\\
&&+8\sum_{k=0}^\infty \left(k+1\right)^{-2s}R^{2s}\left(BR^2s-2\alpha BR^2s(1+s)(k+1)^{-2/3}+BR^2\frac{s}{4}\left(4+BR^2(1+2s)\right)(k+1)^{-1}\right).
\end{eqnarray*}

The first sum is an analytic function for $\mathfrak{R} s>0$. The second sum is evaluated for $\mathfrak{R}s>1/2$ and analytically continued to $s\to 0$: 
\begin{eqnarray*}
&&\sum_{k=0}^\infty \left(k+1\right)^{-2s}R^{2s}\left(BR^2s-2\alpha BR^2s(1+s)(k+1)^{-2/3}+BR^2\frac{s}{4}\left(4+BR^2(1+2s)\right)(k+1)^{-1}\right)\\
&&=BR^{2+2s}s\left(\zeta(2s)-2\alpha(1+s)\zeta(2/3+2s)+\zeta(1+2s)\right)+B^2R^{4+2s}\frac{s}{4}(1+2s)\zeta(1+2s).
\end{eqnarray*}
Finally, we obtain
\begin{eqnarray*}
\delta U^{\textrm{gl}(0)}(B,R)&=&-\frac{1}{2}\left.\frac{d}{ds}\zeta^{\textrm{gl}(0)}(s)\right|_{s=0}\\
&=&-4\sum_{k=0}^\infty \left[-(k+1)\log\frac{\lambda_{\uparrow k\frac{k}{2}0}^2(B,R)}{\lambda_{k0}^2(0,R)}-BR^2\left(1-2\alpha (k+1)^{-2/3}+(k+1)^{-1}\right)-\frac{B^2 R^4}{4(k+1)}\right]\\
&&-2BR^2\left[-1+2\gamma+2\log R-4\alpha\zeta\left(\frac23\right)\right]-B^2R^4(1+\gamma+\log R).
\end{eqnarray*}

\subsection{Contribution of quasi-zero modes with the effective ``mass''}
If one includes the effective ``mass'' $\varkappa$ for quasi-zero modes in the considerations of the previous section, the formulas become
\begin{eqnarray}
\nonumber
&&\delta U_\varkappa^{\textrm{gl}(0)}(B,R)=\frac{1}{2}\textrm{Tr} \log\frac{\lambda_{\uparrow k\frac{k}{2}0}^2(B,R)+\varkappa(BR^2)B}{\lambda_{k0}^2(0,R)}=\left.-\frac{1}{2}\frac{d}{ds}\zeta^{\textrm{gl}(0)}(s)\right|_{s=0},\\
\label{zeta_quasi-zero_gl_with_kappa}
&&\zeta_\varkappa^{\textrm{gl}(0)}(s)=2\mathrm{Tr}\sum_{k=0}^\infty (k+1) \left( \left(\lambda_{\uparrow k\frac{k}{2}0}^2(B,R)+\varkappa(BR^2)B\right)^{-s}-\lambda_{k0}^{-2s}(0,R) \right).
\end{eqnarray}
In analogy to the previous section,
\begin{eqnarray*}
&&\zeta_\varkappa^{\textrm{gl}(0)}(s)=\\
&&=8\sum_{k=0}^\infty \left[(k+1) \left( \left(\lambda_{\uparrow k\frac{k}{2}0}^2(B,R)+\varkappa(BR^2)B\right)^{-s}-\lambda_{k0}^{-2s}(0,R) \right)\right.\\
&&\left.-\left(k+1\right)^{-2s}R^{2s}\left(BR^2s-2\alpha BR^2s(1+s)(k+1)^{-2/3}+BR^2\frac{s}{4}\left(4+BR^2(1+2s)-4\varkappa\left(BR^2\right)\right)(k+1)^{-1}\right)\right]\\
&&+8\sum_{k=0}^\infty \left(k+1\right)^{-2s}R^{2s}\left(BR^2s-2\alpha BR^2s(1+s)(k+1)^{-2/3}+BR^2\frac{s}{4}\left(4+BR^2(1+2s)-4\varkappa\left(BR^2\right)\right)(k+1)^{-1}\right)
\end{eqnarray*}

The first sum is an analytic function for $\mathfrak{R} s>0$. The second sum is evaluated for $\mathfrak{R}s>1/2$ and analytically continued to $s\to 0$: 
\begin{eqnarray*}
&&\sum_{k=0}^\infty \left(k+1\right)^{-2s}R^{2s}\left(BR^2s-2\alpha BR^2s(1+s)(k+1)^{-2/3}+BR^2\frac{s}{4}\left(4+BR^2(1+2s)-4\varkappa\left(BR^2\right)\right)(k+1)^{-1}\right)\\
&&=BR^{2+2s}s\left(\zeta(2s)-2\alpha(1+s)\zeta(2/3+2s)+\left(1-\varkappa\left(BR^2\right)\right)\zeta(1+2s)\right)+B^2R^{4+2s}\frac{s}{4}(1+2s)\zeta(1+2s).
\end{eqnarray*}
Finally,
\begin{eqnarray*}
\delta U_\varkappa^{\textrm{gl}(0)}(B,R)&=&\left.-\frac{1}{2}\frac{d}{ds}\zeta_\varkappa^{\textrm{gl}(0)}(s)\right|_{s=0}\\
&=&-4\sum_{k=0}^\infty \left[-(k+1)\log\frac{\lambda_{\uparrow k\frac{k}{2}0}^2(B,R)+\varkappa(BR^2)B}{\lambda_{k0}^2(0,R)}\right.\\
&&\left.-BR^2\left(1-2\alpha (k+1)^{-2/3}+\left(1-\varkappa\left(BR^2\right)\right)(k+1)^{-1}\right)-\frac{B^2 R^4}{4(k+1)}\right]\\
&&-2BR^2\left[-1+2\left(\gamma+\log R\right)\left(1-\varkappa\left(BR^2\right)\right)-4\alpha\zeta\left(\frac23\right)\right]-B^2R^4(1+\gamma+\log R).
\end{eqnarray*}
\subsection{Contribution of all eigenmodes with the effective ``mass'' for quasi-zero modes}
The desired zeta function corresponding to one-loop correction with the effective ``mass'' for quasi-zero modes is written as
\begin{equation*}
\zeta_\varkappa^\mathrm{gl}(s)=\zeta^\mathrm{gl}(s)-\zeta^{\textrm{gl}(0)}(s)+\zeta_\varkappa^{\textrm{gl}(0)}(s),
\end{equation*}
where zeta functions in right-hand side are given by Eqs.~\eqref{zeta_gl},\eqref{zeta_quasi-zero_gl} and~\eqref{zeta_quasi-zero_gl_with_kappa}. The corresponding effective potential is given by
\begin{eqnarray*}
\delta U_\varkappa^\mathrm{gl}(B,R)&=&-\frac{1}{2}\left.\frac{d}{ds}\zeta_\varkappa^\text{gl}(s)\right|_{s=0}\\
&=&4\sum_{k=1}^\infty k \left[\sum_{m=-\frac{k-1}{2}}^\frac{k-1}{2}\Psi^\text{gl}(k,m,0,\frac{\sqrt{3}B}{2},R)-\frac{3}{4}B^2R^4\frac{1}{24}\left(1-\frac{1}{k}-\frac{5}{k^2}\right)\right]
\\
&&-\frac{B^2R^4}{48}(31+30\gamma-30\log 2+30\log R)+\frac{3B^4R^8}{5120}\\
&&+4\sum_{k=0}^\infty \left[(k+1)\log\frac{\lambda_{\uparrow k\frac{k}{2}0}^2(B,R)+\varkappa(BR^2)B}{\lambda_{k\frac{k}{2}0}^2(B,R)}+BR^2\varkappa\left(BR^2\right)(k+1)^{-1}\right]
+4BR^2\varkappa(BR^2)(\gamma+\log R).
\end{eqnarray*}
Thus obtained effective potential is not an even function of $B$. To restore invariance under $B\to -B$, one adds term $\varkappa$ to contribution of modes $\lambda_{\downarrow k,-\frac{k}{2},0}^2(B,R)$ (these modes become quasi-zero when field $B$ changes direction to the opposite). The function $\varkappa(z)$ should also be even function (possibly constant).
After these steps one obtains formula~\eqref{gluon_effpot_mass} for the effective potential.

\section{Quark eigenmodes\label{appendix_quark}}
The equation for eigenvalues of quarks field in the presence of homogeneous (anti-)self-dual gluon field reads
\begin{equation}\label{the_equation}
\slashed{D} \psi=\lambda\psi.
\end{equation}
Here
\begin{eqnarray*}
&&D_\mu=\partial_\mu-i\hat B_\mu,\quad \hat B_\mu=\hat n B_\mu,\quad B_\mu=-\frac{1}{2} B_{\mu\nu}x_\nu,\quad B_{\mu\nu}=\pm \frac{1}{2}\varepsilon_{\mu\nu\alpha\beta}B_{\alpha\beta},\\
&&\hat{n}=\cos\xi t^3+\sin\xi t^8,\quad \xi=\frac{\pi}{6}+\frac{\pi}{3}k, k=0,1,\dots,5,
\end{eqnarray*}
$t_i$ are generators of $SU(3)$ in fundamental representation.
The field strength tensor $B_{\mu\nu}$ may be parametrized as
\begin{equation*}
B_{ij}=\varepsilon_{ijk}B_k,\quad B_{i4}=\pm B_i,\quad B=\sqrt{B_1^2+B_2^2+B_3^2},
\end{equation*}
where ``$+$'' stands for self-dual field and ``$-$'' for anti-self-dual field.
We choose anti-Hermitian chiral representation of gamma matrices
\begin{equation*}
\left\{\gamma_\mu,\gamma_\nu\right\}=-2\delta_{\mu\nu},\quad
\gamma_i=
\left(
\begin{matrix}
0&\sigma_i\\
-\sigma_i&0
\end{matrix}
\right),\quad
\gamma_4=i
\left(
\begin{matrix}
0&\mathbf{1}\\
\mathbf{1}&0
\end{matrix}
\right),\quad
\gamma_5=\gamma_1\gamma_2\gamma_3\gamma_4=\left(
\begin{matrix}
\mathbf{1}&0\\
0&\mathbf{-1}
\end{matrix}
\right).
\end{equation*}
It is convenient to introduce projectors
\begin{equation*}
\Sigma_\pm=\frac{1}{2}\left(1\pm\frac{\Sigma_iB_i}{B}\right),\quad P_\pm=\frac{1}{2}(1\pm\gamma_5),
\end{equation*}
where
\begin{equation*}
\Sigma_i=\frac{1}{2}\varepsilon_{ijk}\sigma_{jk},\quad \sigma_{\mu\nu}=-\frac{1}{2}\varepsilon_{\mu\nu\alpha\beta}\sigma_{\alpha\beta}\gamma_5,\quad \sigma_{ij}=\varepsilon_{ijk}\Sigma_k,\quad \sigma_{i4}=-\gamma_5\Sigma_i.
\end{equation*}
For self-dual or anti-self-dual field $B_{\mu\nu}$ one obtains identities
\begin{equation*}
\sigma_{\mu\nu}B_{\mu\nu}=4BP_\mp\left(\Sigma_+-\Sigma_-\right),\quad \gamma_\mu B_{\mu\nu}x_\nu P_\mp=-i\slashed{x}BP_\mp \left(\Sigma_+-\Sigma_-\right),
\end{equation*}
where the upper sign corresponds to self-dual field, the lower sign corresponds to anti-self-dual-field. 
Color matrix $\hat{n}$ is diagonal, and for the sake of simplicity we keep the notation $\hat n$ for its diagonal elements.

Acting on equation~\eqref{the_equation} with the projectors $P_\pm=(1\pm\gamma_5)/2$, one rewrites it as
\begin{eqnarray*}
\slashed{D} P_+\psi&=&\lambda P_-\psi,\\
\slashed{D} P_-\psi&=&\lambda P_+\psi.
\end{eqnarray*}
or
\begin{eqnarray*}
\slashed{D} \psi_+&=&\lambda \psi_-,\\
\slashed{D} \psi_-&=&\lambda \psi_+,
\end{eqnarray*}
where $\psi_\pm=P_\pm\psi$.
Substituting one equation into the other (this is valid if $\lambda \neq 0$), one finds
\begin{eqnarray}
\label{squared_equation}
&&\slashed{D}^2 \psi_\mp=\lambda^2 \psi_\mp,\\
\label{chiral_components_relation}
&&\lambda \psi_\pm= \slashed{D} \psi_\mp,\\
\label{full_solution}
&&\psi=\psi_++\psi_-=\left(\frac{\slashed{D}}{\lambda}+1\right)\psi_\mp.
\end{eqnarray}
Only one chiral component is independent, the other one is found via Eq.~\eqref{chiral_components_relation}. The expression for $\gamma_\mu B_{\mu\nu}x_\nu P_\pm$ ($\gamma_\mu B_{\mu\nu}x_\nu$ originates from $\slashed{D}$) is more complex than $\gamma_\mu B_{\mu\nu}x_\nu P_\mp$, so we find $\psi_-$ from equation~\eqref{squared_equation} in the case of self-dual field and $\psi_+$ in the case of anti-self-dual field. Here and below, if $\pm$ or $\mp$ appears alongside with Dirac operator $\slashed{D}$ or its eigenmode $\psi$, the upper sign should be taken for self-dual field and the lower sign for anti-self-dual field.
$\eta$ is the normal to the surface of a sphere, $\eta^2=1$.

The analogue of total angular momentum in Euclidean space is~\cite{Pais:1954}
\begin{eqnarray*}
J_{\mu\nu}&=&K_{\mu\nu}+S_{\mu\nu},\\
K_{\mu\nu}&=&-i\left(x_\mu\partial_\nu-x_\nu\partial_\mu\right),\quad S_{\mu\nu}=\frac{i}{4}\left[\gamma_\mu,\gamma_\nu\right],\\
J_i^\pm&=&\frac{1}{2}\left(\frac{1}{2}\varepsilon_{ijk}J_{jk}\pm J_{i4}\right)=\frac{1}{2}\left(\frac{1}{2}\varepsilon_{ijk}K_{jk}\pm K_{i4}\right)+\frac{1}{2}\left(\frac{1}{2}\varepsilon_{ijk}S_{jk}\pm S_{i4}\right)=K_i^{\pm}+\frac{1}{2}P_\mp \Sigma_i.
\end{eqnarray*}
Algebra of operators $K_{\mu\nu},S_{\mu\nu},J_{\mu\nu}$ splits into two $so(3)$ algebras
\begin{equation*}
[K_i^+,K_j^+]=i\varepsilon_{ijk}K_k^+,\quad [K_i^-,K_j^-]=i\varepsilon_{ijk}K_k^-,\quad [K_i^+,K_j^-]=0,
\end{equation*}
and analogous relations for $J^{\pm}$. One introduces lowering and raising operators
\begin{eqnarray*}
&&\Sigma^{(\pm)}=\frac{1}{2}\left(\Sigma_1\pm i\Sigma_2\right),\quad K^{(\pm)}=K_1\pm i K_2,\\
&&[K_3,K^{(\pm)}]=\pm K^{(\pm)},\quad [K^{(+)},K^{(-)}]=2K_3.
\end{eqnarray*}

We choose reference frame such that direction of field coincides with $z$ axis.
\begin{equation*}
B_i=\left\{0,0,B\right\}.
\end{equation*}
Thus, equation~\eqref{squared_equation} can be cast into the form
\begin{eqnarray*}
\left\{-\left[\frac{1}{r^3}\partial_r r^3 \partial_r -\frac{4}{r^2}\mathbf{K}^2_{\mp}+2\hat{B}K_{2,1}^z-\frac{1}{4}\hat{B}^2r^2\right]-2\hat{B}\right\} \psi_{\mp+}&=&\lambda^2 \psi_{\mp+},\\
\left\{-\left[\frac{1}{r^3}\partial_r r^3 \partial_r -\frac{4}{r^2}\mathbf{K}^2_{\mp}+2\hat{B}K_{2,1}^z-\frac{1}{4}\hat{B}^2r^2\right]+2\hat{B}\right\} \psi_{\mp-}&=&\lambda^2 \psi_{\mp-},
\end{eqnarray*}
where $c$ in $\psi_{cs}$ stands for chirality, and $s$ for spin.
It is convenient to introduce basis spinors $u_{cs}$
\begin{eqnarray*}
&&P_\pm u_{\pm s}=u_{\pm s},\quad P_\pm u_{\mp s}=0,\quad \Sigma_\pm u_{c\pm}=u_{c\pm},\quad \Sigma_\pm u_{c\mp}=0,\quad u^\dagger_{cs}u_{c's'}=\delta_{cc'}\delta_{ss'},\\
&&\Sigma^{(\mp)}u_{c\pm}=u_{c\mp},\quad \Sigma^{(\pm)}u_{c\pm}=0.
\end{eqnarray*}
The solutions are
\begin{eqnarray}
\label{solution_squared1}
\psi_{\mp+}^{km_1m_2}&=& \left(\frac{\hat{B}r^2}{2}\right)^{k/2} e^{-\hat{B}r^2/4}M\left(\frac{k}{2}+1-m_{2,1}+\frac{-2\hat{B}-\lambda^2}{2\hat{B}},k+2,\frac{\hat{B}r^2}{2}\right)Y_{\frac{k}{2}m_1m_2}(\varphi,\chi,\eta)u_{\mp+},\\
\label{solution_squared2}
\psi_{\mp-}^{km_1m_2}&=& \left(\frac{\hat{B}r^2}{2}\right)^{k/2} e^{-\hat{B}r^2/4}M\left(\frac{k}{2}+1-m_{2,1}+\frac{2\hat{B}-\lambda^2}{2\hat{B}},k+2,\frac{\hat{B}r^2}{2}\right)Y_{\frac{k}{2}m_1m_2}(\varphi,\chi,\eta)u_{\mp-}.
\end{eqnarray}
Where $Y_{\frac{k}{2}m_1m_2}$ are spherical harmonics in four-dimensional Euclidean space~\cite{Pais:1954}.

The solutions of~\eqref{the_equation} may be characterized by eigenvalues of independent operators
$
\mathbf{J}^{\mp 2},\ J_3^{\mp},\ J_3^{\pm},\ L
$ (see~\cite{Carter:1979fe,Shapovalov:2018scb,Breev:2015aca}).
Here
\begin{equation*}
L=i\gamma_\mu \gamma_5 B_{\mu\nu}D_\nu,
\end{equation*}
and upper or lower signs should be taken for self-dual field and anti-self-dual field, correspondingly.
The solutions to the eigenvalue problem~\eqref{the_equation}
\begin{equation*}
\left(\frac{\not \hspace*{-0.3em}D}{\lambda}+1\right)\psi_{c s}^{km_1m_2},
\end{equation*}
where $\psi_{c s}^{km_1m_2}$ ($c=\mp,\ s=\pm$) are solutions of Eq.~\eqref{squared_equation} given by Eqs.~\eqref{solution_squared1} and~\eqref{solution_squared2},
diagonalize all these operators.

Now, we impose bag boundary condition~\eqref{bag_bc}. Substituting Eq.~\eqref{full_solution} in Eq.~\eqref{bag_bc} and acting on it with operators $P_\pm$, we find that boundary condition reduces to
\begin{equation*}
i e^{-i\alpha\gamma_5}\hspace*{-0.2em}\not{\hspace*{-0.2em}\eta}\frac{\not \hspace*{-0.3em}D}{\lambda}\psi_\mp(x)=\psi_\mp(x),\quad x^2=R^2,
\end{equation*}
where $\psi_\mp$ are solutions of Eq.~\eqref{squared_equation}. The boundary condition breaks symmetry associated with generator $L$. In order to satisfy boundary condition, we have to mix eigenstates of operator $L$.

Let us consider the case of self-dual field (the case of anti-self-dual field is obtained via $m_2\to m_1,\alpha\to-\alpha$). To satisfy boundary condition, linear combination of solutions with equal projection of total angular momenta $j_3$ is constructed. At given $k$, there is one solution for maximal value of $j_3=\frac{k}{2}+\frac12$, one solution for minimal value $j_3=-\frac{k}{2}-\frac12$ and two solutions for each intermediate $j_3=-\frac{k}{2}+\frac12,\dots,\frac{k}{2}-\frac{1}{2}$ (for $k>0$). Below we consider these three cases.

\subsection{Intermediate values of $j_3$}
As described above, we consider linear combination
\begin{equation*}
\psi_-=C_1 f(k,m_2+1,r)Y_{\frac{k}{2},m_1,m_2}u_{-+}+C_2 f(k,m_2,r) Y_{\frac{k}{2},m_1,m_2+1}(\varphi,\chi,\eta)u_{--}
\end{equation*}
for
\begin{equation*}
m_2=j_3-\frac{1}{2}=-\frac{k}{2},-\frac{k}{2}+1,\dots,\frac{k}{2}-1,\quad k=1,2,\dots,
\end{equation*}
where we introduced notation
\begin{equation*}
f(k,m,r)=\left(\frac{\hat{B}r^2}{2}\right)^{k/2}e^{-\hat{B}r^2/4}M\left(\frac{k}{2}+1-m+\frac{-\lambda^2}{2\hat{B}},k+2,\frac{\hat{B}r^2}{2}\right).
\end{equation*}
The boundary condition reads
\begin{eqnarray*}
\frac{1}{\lambda}\left\{C_1\left[-\partial_r-\frac{\hat{B}R}{2}+\frac{2}{R}m_2\right]f(k,m_2+1,R)+C_2 \sqrt{\frac{k}{2}\left(\frac{k}{2}+1\right)-m_2(m_2+1)} \frac{2}{R}f(k,m_2,R)\right\}=&&\\
=-ie^{-i\alpha}C_1f(k,m_2+1,R),&&
\\
\frac{1}{\lambda}\left\{C_2\left[-\partial_r+\frac{\hat{B}R}{2}-\frac{2}{R}(m_2+1)\right]f(k,m_2,R)+C_1 \sqrt{\frac{k}{2}\left(\frac{k}{2}+1\right)-m_2(m_2+1)} \frac{2}{R}f(k,m_2+1,R)\right\}=&&\\
=-ie^{-i\alpha} C_2f(k,m_2,R).&&
\end{eqnarray*}
Using identities (see~\cite{Abramowitz:1972}, 13.4.10,13.4.11)
\begin{eqnarray*}
&&aM(a+1,b,z)=aM(a,b,z)+zM'(a,b,z),\\
&&(b-a)M(a-1,b,z)=(b-a-z)M(a,b,z)+zM'(a,b,z),
\end{eqnarray*}
we bring these equations to the form
\begin{eqnarray}
\label{BC_general_up}
\nonumber
&&\frac{C_1}{\lambda}\left[ \left(-\frac{\lambda^2}{2\hat{B}}+ie^{-i\alpha}\frac{\lambda R}{2}\right)M\left(\frac{k}{2}-m_2+\frac{-\lambda^2}{2\hat{B}},k+2,\frac{\hat{B}R^2}{2}\right)
-\right.\\
\nonumber
&&\left.-\left(\frac{k}{2}-m_2+\frac{-\lambda^2}{2\hat{B}}\right) M\left(\frac{k}{2}+1-m_2+\frac{-\lambda^2}{2\hat{B}},k+2,\frac{\hat{B}R^2}{2}\right)\right]+\\
&&+\frac{C_2}{\lambda}\sqrt{\frac{k}{2}\left(\frac{k}{2}+1\right)-m_2(m_2+1)} M\left(\frac{k}{2}+1-m_2+\frac{-\lambda^2}{2\hat{B}},k+2,\frac{\hat{B}R^2}{2}\right)=0,
\\
\label{BC_general_down}
\nonumber
&&\frac{C_2}{\lambda}\left[\left(\frac{\lambda^2}{2\hat{B}}+ie^{-i\alpha}\frac{\lambda R}{2}\right) M\left(\frac{k}{2}+1-m_2+\frac{-\lambda^2}{2\hat{B}},k+2,\frac{\hat{B}R^2}{2}\right)-\right.\\
\nonumber
&&\left. -\left(\frac{k}{2}+1+m_2+\frac{\lambda^2}{2\hat{B}}\right) M\left(\frac{k}{2}-m_2+\frac{-\lambda^2}{2\hat{B}},k+2,\frac{\hat{B}R^2}{2}\right)
\right]+\\
&&+\frac{C_1}{\lambda}\sqrt{\frac{k}{2}\left(\frac{k}{2}+1\right)-m_2(m_2+1)}M\left(\frac{k}{2}-m_2+\frac{-\lambda^2}{2\hat{B}},k+2,\frac{\hat{B}R^2}{2}\right)=0.
\end{eqnarray}
This system is uniform with respect to $C_1,C_2$ and has nontrivial solution if the determinant of the system is zero. The latter requirement leads to equation for eigenvalues:
\begin{eqnarray*}
\label{BC_intermediate}
&&\frac{1}{\lambda^2}\left[ \left(-\frac{\lambda^2}{2\hat{B}}+ie^{-i\alpha}\frac{\lambda R}{2}\right)M\left(\frac{k}{2}+\frac{1}{2}-j_3+\frac{-\lambda^2}{2\hat{B}},k+2,\frac{\hat{B}R^2}{2}\right)
-\right.\\
&&\left.-\left(\frac{k}{2}+\frac{1}{2}-j_3+\frac{-\lambda^2}{2\hat{B}}\right) M\left(\frac{k}{2}+\frac{3}{2}-j_3+\frac{-\lambda^2}{2\hat{B}},k+2,\frac{\hat{B}R^2}{2}\right)\right]
\times
\\
&&\times\left[\left(\frac{\lambda^2}{2\hat{B}}+ie^{-i\alpha}\frac{\lambda R}{2}\right) M\left(\frac{k}{2}+\frac{3}{2}-j_3+\frac{-\lambda^2}{2\hat{B}},k+2,\frac{\hat{B}R^2}{2}\right)-\right.\\
&&\left. -\left(\frac{k}{2}+\frac{1}{2}+j_3+\frac{\lambda^2}{2\hat{B}}\right) M\left(\frac{k}{2}+\frac{1}{2}-j_3+\frac{-\lambda^2}{2\hat{B}},k+2,\frac{\hat{B}R^2}{2}\right)
\right]-\\
&&-\frac{1}{\lambda^2}\left[\frac{k}{2}\left(\frac{k}{2}+1\right)-j_3^2+\frac{1}{4}\right]M\left(\frac{k}{2}+\frac{1}{2}-j_3+\frac{-\lambda^2}{2\hat{B}},k+2,\frac{\hat{B}R^2}{2}\right) M\left(\frac{k}{2}+\frac{3}{2}-j_3+\frac{-\lambda^2}{2\hat{B}},k+2,\frac{\hat{B}R^2}{2}\right)=0.
\end{eqnarray*}
where substitution $m_2\to j_3-\frac{1}{2}$ is made.

\subsection{Maximal value of $j_3$}
In this case there is only one function with $j_3=\frac{k}{2}+\frac{1}{2}$ at given $k=0,1,2,\dots$
\begin{equation*}
\psi_-=C_1 f(k,m_2+1,r)Y_{\frac{k}{2},m_1,m_2}u_{-+},\quad m_2=j_3-\frac{1}{2}=\frac{k}{2}.
\end{equation*}
Boundary condition in this case can be obtained from Eq.~\eqref{BC_general_up} with $C_2=0$. After  simplification we find
\begin{equation*}
\label{BC_max}
M\left(-\frac{\lambda^2}{2\hat{B}},k+2,\frac{\hat{B}R^2}{2}\right)
-ie^{i\alpha}\frac{\lambda R}{2(k+2)} M\left(1-\frac{\lambda^2}{2\hat{B}},k+3,\frac{\hat{B}R^2}{2}\right)=0.
\end{equation*}

\subsection{Minimal value of $j_3$}
There is also only one function with $j_3=-\frac{k}{2}-\frac{1}{2}$  at given $k=0,1,2,\dots$
\begin{equation*}
\psi_-=C_2 f(k,m_2,r) Y_{\frac{k}{2},m_1,m_2+1}(\varphi,\chi,\eta)u_{--},\quad m_2=j_3-\frac{1}{2}=-\frac{k}{2}-1.
\end{equation*}
Boundary condition can be obtained from Eq.~\eqref{BC_general_down} with $C_1=0$. After simplification and Kummer transformation (see~\cite{Abramowitz:1972}, 13.1.27)
\begin{equation*}
M(a,b,z)=e^z M(b-a,b,-z).
\end{equation*}
we obtain
\begin{equation*}
\label{BC_min}
M\left(\frac{\lambda^2}{2\hat{B}},k+2,-\frac{\hat{B}R^2}{2}\right)-ie^{i\alpha}\frac{\lambda R}{2(k+2)} M\left(1+\frac{\lambda^2}{2\hat{B}},k+3,-\frac{\hat{B}R^2}{2}\right)=0.
\end{equation*}

\section{Zeta function for quark fields\label{appendix_quark_zeta}}
We put chiral angle $\alpha=\pi/2$, so the Dirac operator is Hermitian and eigenvalues are real. Zeta function $\zeta(s)$ of fermions is split into two parts  
\begin{eqnarray*}
&&\zeta^\mathrm{q}(s)=\cos(\pi s)\zeta_{\slashed{D}^2}\left(\frac{s}{2}\right)-i\sin(\pi s)\eta(s),\\
&&\zeta_{\slashed{D}^2}(s)=\mathrm{Tr}\sum_{k,j_3,n}(k+1) \left(\frac{1}{\left|\lambda_{kj_3n}(\hat{B},R)\right|^{2s}}-
\frac{1}{\left|\lambda_{kj_3n}(0,R)\right|^{2s}}\right),\\
&&\eta(s)=\mathrm{Tr}\sum_{k,j_3,n}(k+1) \left(\frac{\mathrm{sign}{\lambda_{kj_3n}(\hat{B},R)}}{\left|\lambda_{kj_3n}(\hat{B},R)\right|^s}-
\frac{\mathrm{sign}{\lambda_{kj_3n}(0,R)}}{\left|\lambda_{kj_3n}(0,R)\right|^s}\right).
\end{eqnarray*}
Eigenvalues $\lambda_{kjn}$ are found from equations $A(\lambda,k,j_3,\hat{B},R|\alpha)=0$.
The set of eigenvalues for $j_3=\pm\frac{k+1}{2}$ is determined by the equations
\begin{eqnarray*}
&&A(\lambda,k,\frac{k+1}{2},\hat{B},R)=M\left(-\frac{\lambda^2}{2\hat{B}},k+2,\frac{\hat{B}R^2}{2}\right)-\frac{iR\lambda}{2(k+2)}M\left(1-\frac{\lambda^2}{2\hat{B}},k+3,\frac{\hat{B}R^2}{2}\right)=0,\\
&&A(\lambda,k,-\frac{k+1}{2},\hat{B},R)=M\left(\frac{\lambda^2}{2\hat{B}},k+2,-\frac{\hat{B}R^2}{2}\right)-\frac{iR\lambda}{2(k+2)}M\left(1+\frac{\lambda^2}{2\hat{B}},k+3,-\frac{\hat{B}R^2}{2}\right)=0.
\end{eqnarray*}
At $B=0$ the equations take the form
\begin{equation*}
(k+1)!\left(\frac{\lambda R}{2}\right)^{-k-1}\left[J_{k+1}(\lambda R)-iJ_{k+2}(\lambda R)\right]=0.
\end{equation*}

The eigenvalues for $j_3=-\frac{k-1}{2},\dots,\frac{k-1}{2},k\geqslant 1$ are found from equation
\begin{eqnarray*}
&&A(\lambda,k,j_3,\hat{B},R)=\exp\left(-\frac{\hat{B}R^2}{2}\right)\frac{1}{\lambda^2}\left[ i\frac{\lambda R}{2}M\left(\frac{k}{2}+\frac{1}{2}-j_3+\frac{-\lambda^2}{2\hat{B}},k+2,\frac{\hat{B}R^2}{2}\right)
-\right.\\
&&\left.-\left(\frac{k}{2}+\frac{1}{2}-j_3\right) M\left(\frac{k}{2}+\frac{3}{2}-j_3+\frac{-\lambda^2}{2\hat{B}},k+2,\frac{\hat{B}R^2}{2}\right)+\frac{\lambda^2R^2}{4(k+2)}M\left(\frac{k}{2}+\frac{3}{2}-j_3+\frac{-\lambda^2}{2\hat{B}},k+3,\frac{\hat{B}R^2}{2}\right)\right]
\times
\\
&&\times\left[i\frac{\lambda R}{2} M\left(\frac{k}{2}+\frac{3}{2}-j_3+\frac{-\lambda^2}{2\hat{B}},k+2,\frac{\hat{B}R^2}{2}\right)\right.\\
&&\left. -\left(\frac{k}{2}+\frac{1}{2}+j_3\right) M\left(\frac{k}{2}+\frac{1}{2}-j_3+\frac{-\lambda^2}{2\hat{B}},k+2,\frac{\hat{B}R^2}{2}\right)+\frac{\lambda^2R^2}{4(k+2)}M\left(\frac{k}{2}+\frac{3}{2}-j_3+\frac{-\lambda^2}{2\hat{B}},k+3,\frac{\hat{B}R^2}{2}\right)
\right]\\
&&-\exp\left(-\frac{\hat{B}R^2}{2}\right)\frac{1}{\lambda^2}\left[\frac{k}{2}\left(\frac{k}{2}+1\right)-j_3^2+\frac{1}{4}\right]\times\\
&&\times M\left(\frac{k}{2}+\frac{1}{2}-j_3+\frac{-\lambda^2}{2\hat{B}},k+2,\frac{\hat{B}R^2}{2}\right) M\left(\frac{k}{2}+\frac{3}{2}-j_3+\frac{-\lambda^2}{2\hat{B}},k+2,\frac{\hat{B}R^2}{2}\right)=0.
\end{eqnarray*}
Note the factor $\exp\left(-\frac{\hat{B}R^2}{2}\right)$ which does not affect the solutions of the equation, but makes the equation invariant with respect to $j_3\to -j_3,\ B\to -B$ (via Kummer transformation). So, the decomposition of $\zeta(s)$ and the free energy will contain only even powers of $B$. At $B\to 0$ this equation transforms to
\begin{eqnarray*}
&&A(\lambda,k,j_3,0,R|\alpha)=\frac{1}{\lambda^2}(k+1)!\left(\frac{\lambda R}{2}\right)^{-k-1}\left[ i\frac{\lambda R}{2}J_{k+1}(\lambda R)
-\left(\frac{k}{2}+\frac{1}{2}-j\right) J_{k+1}(\lambda R)+\frac{\lambda R}{2} J_{k+2}(\lambda R)\right]
\times
\\
&&\times (k+1)!\left(\frac{\lambda R}{2}\right)^{-k-1}\left[i\frac{\lambda R}{2} J_{k+1}(\lambda R) -\left(\frac{k}{2}+\frac{1}{2}+j\right) J_{k+1}(\lambda R) + \frac{\lambda R}{2} J_{k+2}(\lambda R)
\right]\\
&&-\frac{1}{\lambda^2}\left[\frac{k}{2}\left(\frac{k}{2}+1\right)-j^2+\frac{1}{4}\right]\left[ (k+1)!\left(\frac{\lambda R}{2}\right)^{-k-1} J_{k+1}(\lambda R)\right]^2 =0.
\end{eqnarray*}

It follows that
\begin{equation*}
\zeta_{\slashed{D}^2}(s)=\zeta_{\slashed{D}^2}^{(1)}(s)+\zeta_{\slashed{D}^2}^{(2)}(s),
\end{equation*}
where the first term summarizes contributions of the eigenvalues with maximal and minimal projections of total angular momentum $j_3=\pm\frac{k+1}{2}$, and the second term with intermediate values $j_3=-\frac{k-1}{2},\dots,\frac{k-1}{2}$.
It is easily seen that
\begin{equation*}
A(\lambda,k,j_3,\hat{B},R)=A(\lambda,k,-j_3,-\hat{B},R),
\end{equation*}
so the invariance of the spectrum under transformation $B\to -B$ is manifest. 

\subsection{Contribution of maximal and minimal $j_3$}
The following considerations are analogous to Appendix~\ref{appendix_ghost_gluon}. We start with the representation
\begin{eqnarray*}
\zeta_{\slashed{D}^2}^{(1)}(s)&=&\mathrm{Tr}\left\{\frac{\sin\pi s}{\pi}\sum_{k=0}^\infty \sum_{j_3=\pm\frac{k+1}{2}} (k+1) \int_0^\infty \frac{dt}{t^{2s}}\frac{d}{dt}\Psi^\mathrm{q}(k+1,j_3,t,\hat{B},R)\right\}\\
&=&\mathrm{Tr}\left\{
\frac{\sin\pi s}{\pi}\sum_{k=1}^\infty k^{1-2s} \int_0^\infty \frac{dt}{t^{2s}}\frac{d}{dt}\left[\sum_{j_3=\pm\frac{k}{2}}\Psi^\mathrm{q}(k,j_3,kt,\hat{B},R)-\sum_{i=1}^2\frac{u_i^\mathrm{q}(t,\hat{B},R)}{k^i}\right]\right.\\
&&+\left.\frac{\sin\pi s}{\pi}\sum_{k=1}^\infty k^{1-2s} \int_0^\infty \frac{dt}{t^{2s}}\frac{d}{dt}+\sum_{i=1}^2\frac{u_i^\mathrm{q}(t,\hat{B},R)}{k^i}\right\}.
\end{eqnarray*}
The functions $u_i^\mathrm{q}$ are given by
\begin{eqnarray*}
u_1^\mathrm{q}(t,\hat{B},R)&=&\hat{B}^2R^4\frac{-8-4R^2t^2+R^4t^4+8\sqrt{1+R^2t^2}}{6R^4t^4\sqrt{1+R^2t^2}},\\
u_2^\mathrm{q}(t,\hat{B},R)&=&\hat{B}^2R^4\frac{-8-12R^2t^2-3R^4t^4+8\sqrt{1+R^2t^2}+8R^2t^2\sqrt{1+R^2t^2}}{4R^4t^4\left(1+R^2t^2\right)^{\frac32}}.
\end{eqnarray*}
Evaluating the sums and integrals, one finds
\begin{equation*}
\sum_{k=1}^\infty k^{1-2s} \int_0^\infty \frac{dt}{t^{2s}}\frac{d}{dt}\sum_{i=1}^2 \frac{u_i^{\mathrm{q}}(t,\hat{B},R)}{k^i}=
\hat{B}^2R^{4+2s}\left\{\zeta(2s)\frac{s\Gamma(1-s)\Gamma\left(s+\frac{1}{2}\right)}{2\sqrt{\pi}(s+2)}-
\zeta(1+2s)\frac{s\Gamma(1-s)\Gamma\left(s+\frac{3}{2}\right)}{2\sqrt{\pi}(s+2)}
\right\}.
\end{equation*}
Expansion of the counterterm in powers of $s$ is
\begin{equation*}
\frac{\sin\pi s}{\pi}\sum_{k=1}^\infty k^{1-2s} \int_0^\infty \frac{dt}{t^{2s}}\frac{d}{dt}\sum_{i=0}^2 \frac{u_i^{\mathrm{q}}(t,\hat{B},R)}{k^i}=-\frac{1}{16}\hat{B}^2R^4s+O(s^2),
\end{equation*}
and the corresponding contribution to the effective potential reads
\begin{equation*}
\frac{1}{2} \left.\frac{d}{ds} \zeta^{(1)}_{\slashed{D}^2}\left(s\right)\right|_{s=0}=-\frac{1}{32}\mathrm{Tr}\hat{B}^2R^4=-\frac{1}{64}B^2R^4.
\end{equation*}

\subsection{Contribution of intermediate $j_3$}
The starting expression in this case is
\begin{eqnarray*}
\zeta_{\slashed{D}^2}^{(2)}(s)&=&\mathrm{Tr}\left\{\frac{\sin\pi s}{\pi} \sum_{k=1}^\infty \sum_{j_3=-\frac{k-1}{2}}^{\frac{k-1}{2}} (k+1) \int_0^\infty \frac{dt}{t^{2s}}\frac{d}{dt}\Psi^\mathrm{q}(k+1,j_3,t,\hat{B},R)\right\}\\
&=&\mathrm{Tr}\left\{\frac{\sin\pi s}{\pi}\sum_{k=1}^\infty (k+1)k^{-2s} \int_0^\infty \frac{dt}{t^{2s}}\frac{d}{dt}\left[\sum_{j_3=-\frac{k-1}{2}}^\frac{k-1}{2}\Psi^\mathrm{q}(k+1,j_3,kt ,\hat{B},R)-\sum_{i=0}^2 \frac{u_i^\mathrm{q}(t,\hat{B},R)}{k^i}\right]\right.
\\
&&\left.
+\frac{\sin\pi s}{\pi}\sum_{k=1}^\infty (k+1) k^{-2s} \int_0^\infty \frac{dt}{t^{2s}}\frac{d}{dt}\sum_{i=0}^2 \frac{u_i^\mathrm{q}(t,\hat{B},R)}{k^i}\right\},
\end{eqnarray*}
where
\begin{eqnarray*}
u_0^\mathrm{q}(t,B,R)&=&\hat{B}^2R^4\frac{R^4t^4-2R^2t^2-4+4\left(1+R^2t^2\right)^{1/2}}{6R^4 t^4\left(1+R^2t^2\right)^{1/2}},\\
u_1^\mathrm{q}(t,\hat{B},R)&=& \hat{B}^2 R^4\frac{-3R^4t^4-8+8\left(1+R^2t^2\right)^{1/2}-12R^2t^2+8R^2t^2\left(1+R^2t^2\right)^{1/2}}{6t^4\left(1+R^2t^2\right)^\frac{3}{2}},\\
u_2^\mathrm{q}(t,\hat{B},R)&=&\frac{\hat{B}^2R^4}{48\left(1+R^2t^2\right)^{1/2}}\left[-\frac{32}{R^2t^2}+\frac{R^2t^2(8+13R^2t^2)}{(1+R^2t^2)^3}-64\frac{1-\left(1+R^2t^2\right)^{1/2}}{R^4t^4}\right]\\
&&+\frac{\hat{B}^4R^8}{480\left(1+R^2t^2\right)^{1/2}}
\left[-\frac{128}{R^6t^6}+\frac{32}{R^4t^4}-\frac{16}{R^2t^2}+\frac{10+13R^2t^2}{(1+R^2t^2)^2}-256\frac{1-\left(1+R^2t^2\right)^{1/2}}{R^8t^8}\right].
\end{eqnarray*}
We find
\begin{eqnarray*}
&&\sum_{k=1}^\infty (k+1)k^{-2s} \int_0^\infty \frac{dt}{t^{2s}}\frac{d}{dt}\sum_{i=0}^2 \frac{u_i^{\mathrm{q}}(t,\hat{B},R)}{k^i}=\\
&&\hat{B}^2R^{4+2s}\left\{\vphantom{\frac11}\right.\zeta(-1+2s)\frac{-\Gamma(2-s)\Gamma(1/2+s)}{6\sqrt{\pi}(2+s)}+
\zeta(2s)\frac{-(1+2s^2)\Gamma(1-s)\Gamma(1/2+s)}{6\sqrt{\pi}(2+s)}+\\
&&\left.
\zeta(1+2s)\frac{(12+s(4-s)(1+s))\Gamma(1-s)\Gamma(3/2+s)}{18\sqrt{\pi}(2+s)}
\right\}-
\hat{B}^4R^{8+2s}\zeta(1+2s)\frac{s\Gamma(2-s)\Gamma(3/2+s)}{240\sqrt{\pi}(4+s)}.
\end{eqnarray*}
The small-$s$ expansion is
\begin{eqnarray*}
&&\frac{\sin\pi s}{\pi}\sum_{k=1}^\infty (k+1)k^{-2s} \int_0^\infty \frac{dt}{t^{2s}}\frac{d}{dt}\sum_{i=0}^2 \frac{u_i^{\mathrm{q}}(t,\hat{B},R)}{k^i}\\
&&=\hat{B}^2R^4\left[\frac{1}{12}+\frac{1}{144}(29+24\gamma-24\log 2+4\pi^2+24\log R)s\right]+
\hat{B}^4R^8\left[-\frac{1}{1920}s\right]+O(s^2).
\end{eqnarray*}
And we find second part of the effective potential
\begin{eqnarray*}
&&\frac12\left.\frac{d}{ds} \zeta_{\slashed{D}^2}^{(2)}\left(s\right)\right|_{s=0}=-\frac{1}{2}\mathrm{Tr}\sum_{k=1}^\infty (k+1) \left[\sum_{j_3=-\frac{k-1}{2}}^\frac{k-1}{2}\Psi^\text{q}(k,j_3,0,\hat{B},R)-\hat{B}^2R^4\frac{1}{12}\left(1-\frac{2}{k^2}\right)\right]\\
&&+\frac{B^2R^4}{576}(29+24\gamma-24\log 2+4\pi^2+24\log R)-
\frac{B^4R^8}{30720}.
\end{eqnarray*}
Combining together two contributions $\zeta_{\slashed{D}^2}^{(1)}$ and $\zeta_{\slashed{D}^2}^{(2)}$ considered above, one obtains Eq.~\eqref{quarks_effpot}.

\section{Connection between Kummer and Bessel functions\label{appendix_Kummer_functions}}
We find the desired limit from definition of Kummer function via series (see~\cite{Abramowitz:1972}, 13.1.2)
\begin{equation*}
\lim_{z\to 0} M\left(a+\frac{b}{z},c,dz\right)=\lim_{z\to 0} \sum_{n=0}^\infty \frac{\left(a+\frac{b}{z}\right)_n(dz)^n}{(c)_n n!}=\sum_{n=0}^\infty \frac{(bd)^n}{(c)_n n!}={_0F_1}(c,bd).
\end{equation*}
Here $(c)_n$ is Pochhammer symbol:
\begin{equation*}
(c)_0=1,\quad (c)_n=c(c+1)\cdots(c+n-1).
\end{equation*}
Comparing series expansions of Bessel functions (\cite{Abramowitz:1972}, 9.1.10, 9.6.10) and hypergeometric function ${}_0F_1$, we find
\begin{eqnarray*}
{_0F_1}(a,z)&=&\sum_{k=0}^\infty \frac{z^k \Gamma(a)}{k! \Gamma(a+k)}=(a-1)!\left(\sqrt{z}\right)^{1-a} \left(\sqrt{z}\right)^{a-1} \sum_{k=0}^\infty \frac{(\sqrt{z})^{2k} }{k! \Gamma(a-1+k+1)}=(a-1)! \left(\sqrt{z}\right)^{1-a} I_{a-1}(2\sqrt{z}),\\
{_0F_1}(a,-z)&=&\sum_{k=0}^\infty \frac{(-z)^k \Gamma(a)}{k! \Gamma(a+k)}=(a-1)!\left(\sqrt{z}\right)^{1-a} \left(\sqrt{z}\right)^{a-1} \sum_{k=0}^\infty \frac{(-1)^k(\sqrt{z})^{2k} }{k! \Gamma(a-1+k+1)}=(a-1)! \left(\sqrt{z}\right)^{1-a} J_{a-1}(2\sqrt{z}).
\end{eqnarray*}
for $z\geqslant 0$.

\section{Asymptotic expansion for quasi-zero eigenvalues\label{appendix_quasi-zero_expansion}}

Let $x$ stands for the smallest solution of the equation.
The equation
\begin{equation}
\label{appendix_quasi-zero_equation}
M\left(-\frac{x(z)^2}{2z},k+2,\frac{z}{2}\right)=0,\quad
k=0,1,2,\dots
\end{equation}
reduces to
\begin{equation*}
\left(\frac{x(0)}{2}\right)^{-k-1}J_{k+1}(x(0))=0,\quad k=0,1,\dots
\end{equation*}
at $z\to 0$.
The solution can be sought in the form of a power series in $z$
\begin{equation}
\label{x_power_series}
x(z)=\sum_{i=0}^\infty \frac{z^i}{i!}\frac{\partial^i}{\partial z^i}x^{(i)}(0).
\end{equation}
The derivatives $x^{(i)}(0)$ can be found successively from the equations
\begin{equation*}
\frac{\partial^i}{\partial z^i} \left.M\left(-\frac{x(z)^2}{2z},k+2,\frac{z}{2}\right)\right|_{z=0}=0.
\end{equation*}
We find
\begin{eqnarray*}
\left.\frac{dx(z)}{dz}\right|_{z=0}&=&-\frac{k+2}{2x(0)},\\
\left.\frac{d^2x(z)}{dz^2}\right|_{z=0}&=& -\frac{k^2+8k+12-x^2(0)}{12 x^3(0)},\\
\left.\frac{d^3x(z)}{dz^3}\right|_{z=0}&=& -\frac{(k+2)(k^2+8k+12-x^2(0))}{8 x^5(0)},
\end{eqnarray*}
where we used identities (see~\cite{Abramowitz:1972}, 9.1.27)
\begin{eqnarray*}
&&J_{\nu-1}(z)+J_{\nu+1}=\frac{2\nu}{z}J_\nu(z),\\
&&J'_\nu(z)=-J_{\nu+1}(z)+\frac{\nu}{z}J_\nu(z)
\end{eqnarray*}
and the fact that
\begin{equation*}
J_{k+1}(x(0))=0.
\end{equation*}
Next, we use uniform asymptotic expansion of zeros of Bessel functions $J_\nu(z)$ (see~\cite{NIST:DLMF}, Eq.~10.21.vii)
\begin{equation}
\label{bessel_zero_expansion}
\rho_\nu(t)=\nu \sum_{k=0}^\infty\frac{\alpha_k}{\nu^{2k/3}},\quad \theta\left(-2^\frac{1}{3}\alpha\right)=\pi t, \quad
\alpha_0=1,\quad \alpha_1=\alpha,\quad \alpha_2=\frac{3}{10}\alpha^2,\dots
\end{equation}
where $\theta(x)$ is the phase of Airy functions
\begin{equation*}
\theta(z)=\arctan\frac{\mathrm{Ai}(z)}{\mathrm{Bi}(z)}.
\end{equation*}
Quasi-zero solutions of~\eqref{appendix_quasi-zero_equation} at $z\to 0$ become smallest zeroes of Bessel functions that correspond to $t=1$.
With $t=1$ formula~\eqref{bessel_zero_expansion} gives the desired expansion for the first zero of $J_{k+1}(z)$
\begin{equation}
\label{bessel_zero_expansion2}
x(0)=(k+1)\sum_{i=0}^\infty \frac{\alpha_k}{(k+1)^{2i/3}},
\end{equation}
where $\alpha\approx 1.855757$. This series can be once again expanded in powers of $k$.

According to Eq.~\eqref{bessel_zero_expansion2}, $x(0)\sim k$ for $k\gg 1$. One notices that power of leading-asymptotics term of derivative $x^{(i)}(0)$ is smaller for larger $i$
\begin{equation*}
x(0)\sim k, \left.\frac{dx(a)}{da}\right|_{a=0}\sim k^{0},\dots
\end{equation*}
so we need only several terms in series~\eqref{x_power_series} to find asymptotic expansion of $x(z)$ in $k$ up to given order:
\begin{equation}
\label{zero_asymptotic_expansion}
x(z)=k+\alpha k^{1/3}+1+\frac{3\alpha^2}{10}k^{-1/3}+\frac{\alpha}{3}k^{-2/3}+\left(\frac{1}{70}-\frac{\alpha^3}{350}\right)k^{-1}+z\left[-\frac{1}{2}+\frac{\alpha}{2}k^{-2/3}-\frac{1}{2}k^{-1}\right]+O(k^{-4/3}).
\end{equation}

\bibliography{references}

\end{document}